\documentclass[fleqn,10pt]{wlscirep}
\usepackage[utf8]{inputenc}
\usepackage[T1]{fontenc}
\usepackage{bm}

\newcommand{\Nv}{S}
\def\<{\langle}
\def\>{\rangle}

\def\v{\textbf{v}}
\def\y{\textbf{y}}
\def\sx{{\sf x}}
\def\sy{{\sf y}}

\def\f{{\bf f}}
\def\nbsubjects{95}
\def\overlappingprobability{$0.79(1)$}

\title{Subjectivity and complexity of facial attractiveness}

\author[1,*]{Miguel Ib{\'a}{\~n}ez-Berganza}
\author[1]{Ambra Amico}
\author[1,2,3]{Vittorio Loreto}
\affil[1]{Sapienza University of Rome, Physics Department, Piazzale Aldo Moro 2, 00185 Rome, Italy}
\affil[2]{Sony Computer Science Laboratories, Paris, 6, rue Amyot, 75005, Paris, France}
\affil[3]{Complexity Science Hub, Josefst\"adter Strasse 39, A 1080 Vienna, Austria}

\affil[*]{miguel.berganza@roma1.infn.it}

\begin{abstract}
The origin and meaning of facial beauty represent a longstanding puzzle. Despite the profuse literature devoted to facial attractiveness, its very nature, its determinants and the nature of inter-person differences remain controversial issues. Here we tackle such questions proposing a novel experimental approach in which human subjects, instead of {\it rating} natural faces, are allowed to efficiently explore the face-space and ``sculpt'' their favorite variation of a reference facial image. The results reveal that different subjects prefer distinguishable regions of the face-space, highlighting the essential subjectivity of the phenomenon.The different sculpted facial vectors exhibit strong correlations among pairs of facial distances, characterising the underlying universality and complexity of the cognitive processes, and the relative relevance and robustness of the different facial distances.
\end{abstract}
\begin{document}

\flushbottom
\maketitle
%
%
\thispagestyle{empty}


\section*{Introduction \label{sec:intro}}

The notions of body beauty and harmony of proportions have fascinated scholars for centuries. From the ancient Greek canons, a countless number of studies have focused on unfolding what is behind the beauty of the face and the body. Nowadays the notion of facial beauty is a fast expanding field in many different disciplines including developmental psychology, evolutionary biology, sociology, cognitive science and neuroscience~\cite{bzdok2011,hahn2014,laurentini2014,little2014,thornhill1999}. Still, despite a profuse and multi-disciplinary literature, questions like the very nature of facial attractiveness, its determinants, and the origin of inter-subject variability of aesthetic criteria, elude a satisfactory understanding. Here, we revisit the question drawing conclusions based on an empirical approach through which we allow human subjects to ``sculpt'' their favorite facial variations by navigating the so called face-space and converging on specific {\it attractors}, or preferred regions in the face-space. 

The face is the part of the human body from which we infer the most information about others, such as: gender, identity, intentions, emotions, attractiveness, age, or ethnicity \cite{walker2016,little2011,leopold2010}. In particular, looking at a face, we are able to immediately acquire a consistent impression of its attractiveness. Still, we could have a hard time explaining what makes a face attractive to us. As a matter of fact, which variables determine attractiveness and their interactions are still poorly understood issues~\cite{laurentini2014}.  
  
Many works have been devoted to assessing the validity of the {\it natural selection hypothesis}, or beauty as a ``certificate'' of  good phenotypic condition \cite{little2011}. According to this hypothesis, a face is judged on average as attractive according to a set of innate rules typical of the human species, which stand out with respect to other social or individual factors.  Some degree of consensus has, indeed, been reported \cite{maret1985,langlois2000,cunningham1995,bernstein1982,thakerar1979}. Most of these experiments are based on the measurement of correlations among {\it numerical ratings} assigned to a set of natural (or synthetic \cite{perrett1994,perrett1998}) facial images by raters belonging to different cultural groups. 
Much work in this field has also been devoted to assessing the covariation of the perceived beauty of a face  with facial {\it traits} that are believed to signal good phenotypic condition, mainly: facial symmetry, {\it averageness} and secondary sexual traits. After decades of intense research, the role played by these traits is known to be limited: facial beauty seems to be more complex than symmetry \cite{thornhill1999}, averageness \cite{alley1991,perrett1994} and secondary sexual traits \cite{valenzano2006,little2011}. 

Indeed, it has been documented that cultural, between-person and intra-person differences influence attractiveness perception in various ways \cite{little2014}. As a representative example, the link between masculinity and attractiveness in male faces  is subject to significant inter- and intra-subject differences \cite{thornhill1999,little2011,little2014,cunningham1990}. An evolutionary explanation is that  exaggerated masculinity could be perceived as denoting a lack of {some personality facets} such as honesty or expressiveness \cite{perrett1998}. In this context, the so called {\it multiple fitness} or {\it multiple motive model}  \cite{cunningham1995,edler2001,little2014} proposes that attractiveness varies according to a variety of {\it motives}, each one evoking a different abstract attribute of the person whose face is evaluated.  

On the other hand, an impressive amount of work is committed to the automatic facial beauty rating. This is tackled as a supervised inference problem whose training database is composed of natural facial images codified by {\it vectors of facial coordinates} in {\it face-space} \cite{valentine2016,hill2011,laurentini2014}, along with (inter-subject averaged) numerical ratings assigned to them by human subjects, to be inferred. Works differ mainly on the codification of faces in the face-space: from a {\it geometric} face description (2D or 3D spatial coordinates of the {\it facial landmarks}), to a detailed description of the {\it texture} or luminosity degrees of freedom that provide a cue to the facial shape in depth (there also exist {\it holistic} representations, extracting lower-dimensional, non-local information from the facial image according to some criterion (Principal Component {\it eigenfaces} or Gabor filters); or using  richer techniques  which integrate geometric from skin textural and reflectivity characteristics). With the advent of deep hierarchical neural networks, the raw facial image is given as an input to the algorithm, which automatically extracts the putative relevant features in the inference process, although in a hardly accessible way (the {\it black box problem}).
   
The supervised inference of ratings may help to address, albeit indirectly,  the impact of various facial features on attractiveness. 
Although the relative relevance of different features has been discussed in various articles, robust conclusions are lacking \cite{joy2006,schmid2008,pallett2010,fan2012,shen2016,gray2010,laurentini2014,bottino2012}. The results  about the relative relevance of the {\it kind} (geometric, textural and holistic) of facial attributes to attractiveness are controversial as well \cite{laurentini2014,eisenthal2006,mu2013,gan2014,bronstad2008,chen2010}. In any case, the integration of different kinds of variables seems to improve the inference results \cite{eisenthal2006,xu2017}, suggesting that these are complementarily taken into account in the cognitive process of attractiveness assessment.

Facial beauty is, hence, probably not a universal function of a set of few facial properties, as implicitly assumed in many references, but the result of a complex process in which multiple semantic concepts, providing cues to personality facets, are inferred. The literature concerning inference of personality traits indicates that such semantic concepts may be encoded in global combinations of facial features, in a complex way \cite{galantucci2014}. This motivates a study of facial beauty beyond the subject-averaged rating, focusing on the inter-subject heterogeneity and on the global  combinations of various facial features generating such a diversity. 

In summary, the complexity of facial attractiveness perception so far prevented a satisfactory understanding of how attractiveness relates to various facial elements \cite{laurentini2014}, and of the nature of inter-personal differences. In order to make progress, from a methodological point of view it is important to highlight three key factors. (A) The possible mutual influence among geometric, texture and detailed features \cite{adolphs2016}. Even considering the problem in terms of geometric variables only, the possible existence of {\it interactions} or mutual dependencies between different facial components may induce a variety of possible pleasant faces, even for the single subject. (B) The undersampling of the relevant face-space, due to the many different prototypes of facial beauty \cite{eisenthal2006,perrett1994}.
 (C) The subjectivity of the phenomenon, probably hindered by the use of the average numerical beauty ratings. The complexity and richness of the perceptual process, suggested by the multiple-motive hypothesis and by previous work about perception of personality dimensions \cite{oosterhof2008,todorov2011,walker2016,abir2017}, eludes a description in terms of average ratings, a quantity that has already been observed to be inadequate \cite{laurentini2014}.

In light of these considerations, we here address the phenomenon of facial preference through an empirical approach that aims at removing the biases of ratings, focusing instead on the possibility given to human subjects to freely explore a suitably defined face-space. By means of a dedicated software, based on image deformation and genetic algorithms, we focus on inter-subject differences in aesthetic criterion and let several subjects sculpt their favorite variation of a {\it reference portrait}, parametrized by a vector of geometric facial coordinates. We observe how different subjects tend to systematically sculpt facial vectors in different regions of the face-space, which we call {\it attractors}, pointing towards a strong subjectivity in the perception of facial beauty. In addition, the facial vectors sculpted by different subjects exhibit strong correlations for pairs of facial distances, which is a manifestation of the underlying universality and complexity of the cognitive process of facial image discrimination. The correlations contain information regarding the different sources of variability in the dataset of selected vectors. For instance, though a difference between male-female subjects is clearly observed, the largest differences among facial variations, elicited by a principal component analysis, result from criteria that are transversal with respect to the gender only. A third important result concerns the assessment of the robustness of the results with respect to the degrees of freedom  not described in the face-space. Crucially, in our approach, the luminance, texture and detailed degrees of freedom are decoupled from the geometric features defining the face-space, and deliberately kept {\it fixed, and common for all the subjects}. Finally, we observe that the overall experimental results are, interestingly, partially robust and independent of the detailed degrees of freedom (the reference portrait).

The current experimental scheme bypasses the three confounding factors (A-C) mentioned in the precedent paragraph. (A) Uncontrolled sources of biases are absent in our study, since all possible facial variations (given the reference portrait) are described by points in the face-space. (B) In our face-space of reduced dimensionality and unchanged texture degrees of freedom the undersampling is mitigated, making possible an efficient exploration of the face space and allowing for an accurate characterisation of the single-subject attractor. (C) This allow us to fully account for subjectivity: we are able to analyse the differences among different subject’s preferred facial modifications.

\section*{Results \label{sec:results}}

\subsection*{Preferred facial images as extrema in face-space}

\begin{figure}[t!]                        
\begin{center} 
\includegraphics[width=.5\columnwidth]{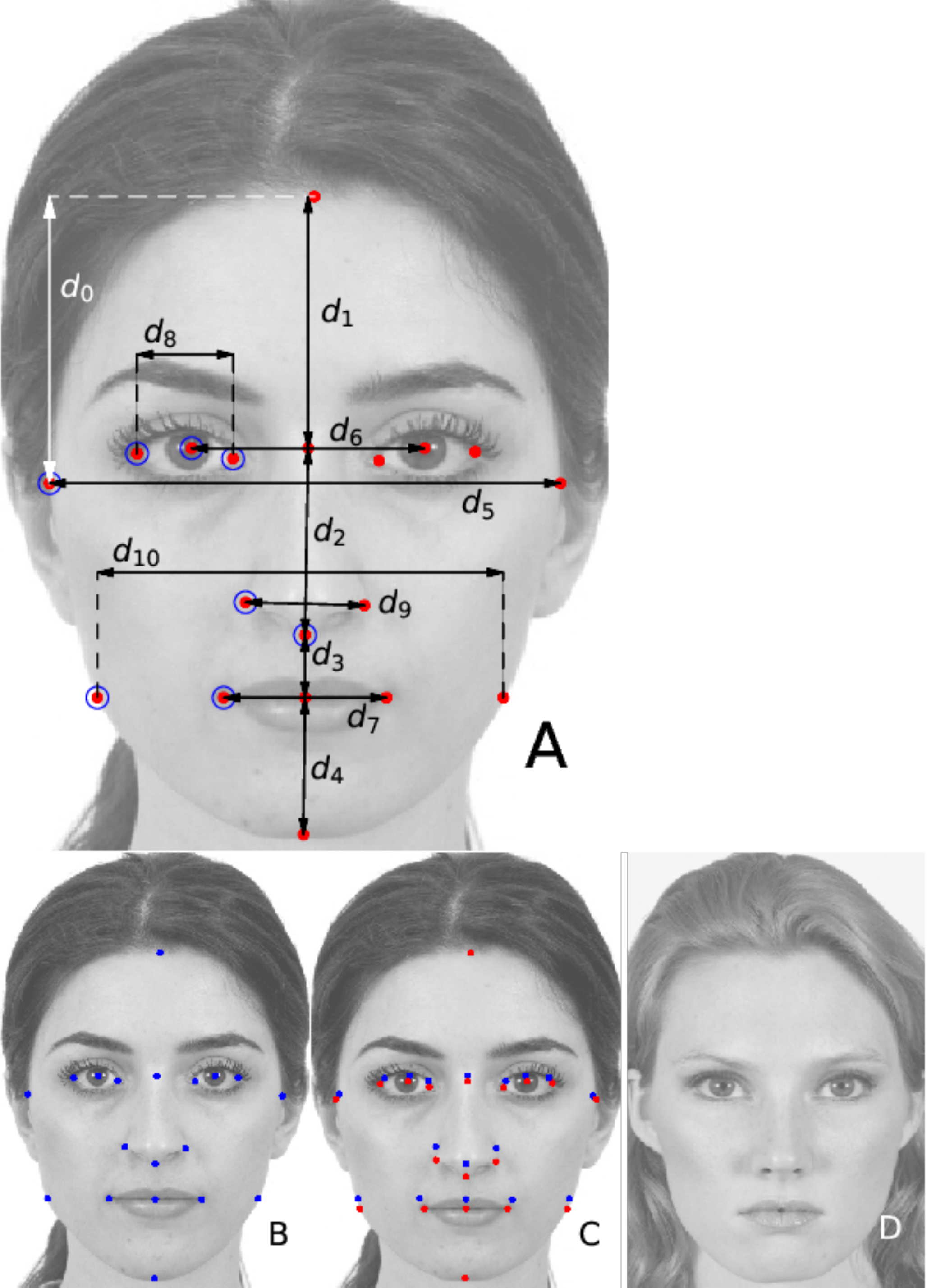}  
\caption{
	{\bf (1-A)} The parameters defining the face space. The red points indicate the {\it landmarks}, $\alpha=1,\ldots,18$, whose 2D varying Cartesian coordinates generate the continuum of face space. The face space points are parametrised in terms of vectors $\bf f$ whose components are the Cartesian coordinates of a set of non-redundant landmarks ${\vec\ell}_\alpha$ (signaled with an empty circle), or in terms of (vertical or horizontal) distances $d_i$ ($i=0,\ldots,10$) among some pairs of landmarks $d_i=|{\sf x}_{\alpha(i)}-{\sf x}_{\beta(i)}|$ or $d_i=|{\sf y}_{\alpha(i)}-{\sf y}_{\beta(i)}|$ (arrows). {\bf (1-B)} Reference portrait RP1 used in experiment E1 along with its corresponding landmarks (in blue). {\bf (1-C)} Image deformation of RP1 according to a given vector of inter-landmark distances $\bf d$: the blue reference portrait landmarks are shifted (leading to the red points) so that their inter-landmark distances are $\bf d$, and the reference image (1-B) is consequently deformed. {\bf (1-D)} Image deformation of the reference portrait RP2 according to the same vector of distances $\bf d$ as in (1-C).
}
\label{fig:facespace}
\end{center}   
\end{figure}

We consider a face-space defined by a set of geometric coordinates illustrated in Fig.~\ref{fig:facespace}-A. A face is parametrized in terms of a set of $10$ non-redundant Cartesian coordinates of $7$ single landmarks $\vec\ell_\alpha=(\sx_\alpha,\sy_\alpha)$ or, alternatively, in terms of a vector of $D=11$ inter-landmark distances ${\bf d}=(d_i)_{i=1}^D$. The face-space vector components $f_i$ are, in this way, either landmark Cartesian coordinates or inter-landmark distances. From a vector of facial coordinates $\bf f$ and a {\it reference facial portrait} corresponding to a real person, we then construct a facial image by a continuous deformation of the reference portrait such that its landmark geometric coordinates acquire the desired value, $\bf f$ (Fig.~\ref{fig:facespace}-B and C). Within a single experiment, the reference portrait (the image texture) is unchanged and only the geometric position of the landmarks can change (for an in-depth explanation see Sec. Methods and the Supplementary Information). 

The aim of the experimental method is to provide a {\it population} of $N$ facial vectors, $\{{\bf f}^{(s,n)}\}_n$, with $n=1,\ldots,N$ and ${\bf f}^{(s,n)}\in \mathbb{R}^D$, for each experimental subject, $s$. Such a population is considered as an empirical sample of the subject's attractor, or the face-space region of his/her preferred modifications of the reference portrait. This means that the subject would probabilistically prefer facial images associated with vectors that are close to the attractor, rather than local fluctuations away from it (for a precise definition see the Supplementary section S2). In our experimental scheme, the subject does not sculpt the population by successive discrimination among faces differing by a single coordinate, which turns out to be an inefficient strategy of face-space exploration, but rather through the interaction with a genetic algorithm (see sections Methods, Supplementary section S3).

\begin{figure}[t!]                        
\begin{center} 
\includegraphics[width=.8\columnwidth]{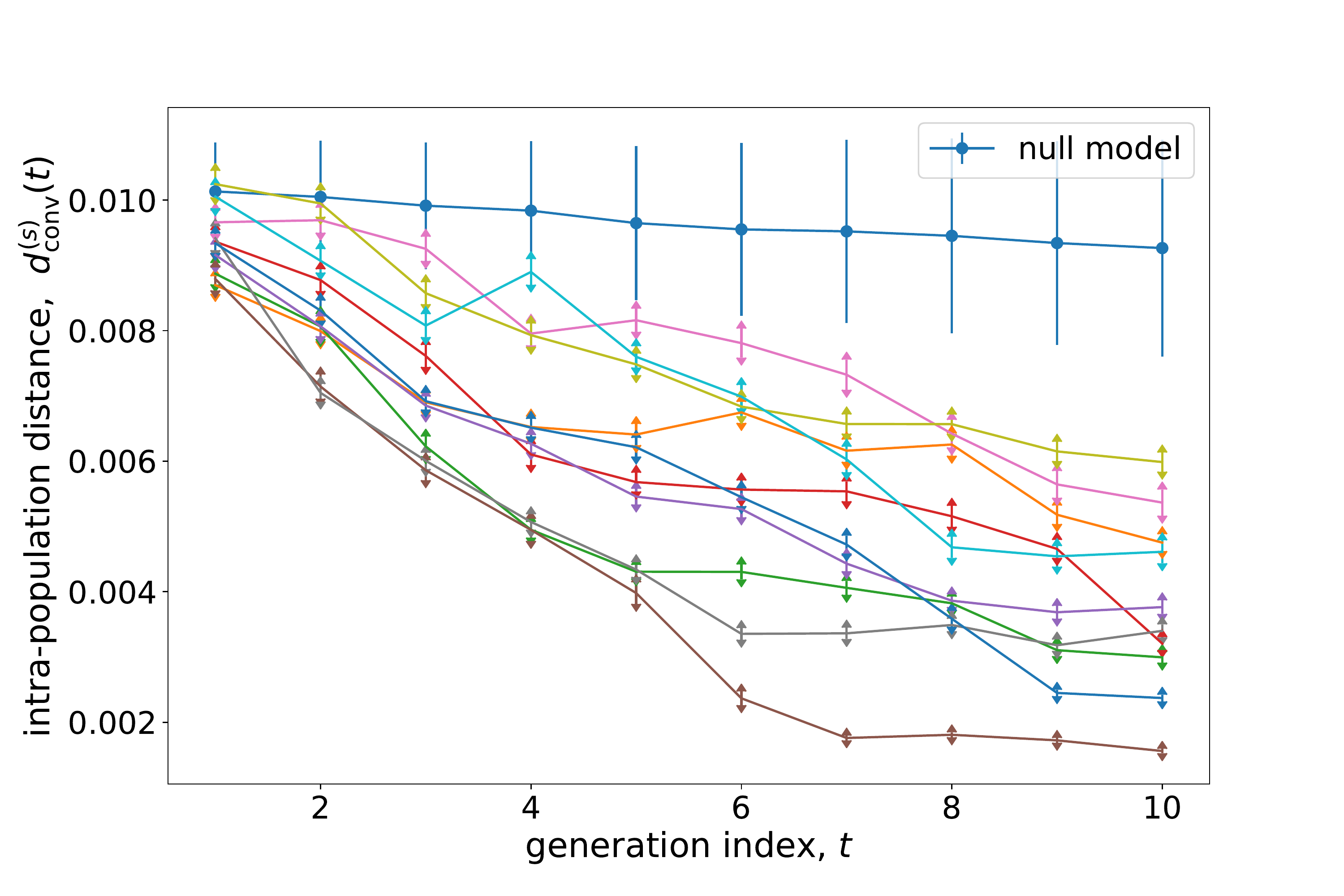}  
	\caption{Intra-population distance of the populations sculpted by different subjects ($s$) as a function of the generation ($t$). The Euclidean metrics in face space  has been used (see Supplementary Sec. S4), although the results are qualitatively equal for other relevant metrics. Each curve corresponds to a different subject (for 10 randomly chosen subjects). The upper curve of joined circles corresponds to the null model genetic experiment, in which the left/right choices are random.}
\label{fig:convergence}
\end{center}   
\end{figure}

In a first experiment (E1), we have let $S_1=\nbsubjects$ subjects sculpt their facial variations of reference portrait RP1 (see ~\ref{fig:facespace}-A). This results in a final population, ${\cal S}_1=\{{\bf f}^{(s,n)}\}_{s=1,n=1}^{S_1,N}$ of $N=28$ facial vectors for each subject. Starting from $N$ initial random facial vectors, the FACEXPLORE software generates pairs of facial images that are presented to the subject, who selects the one that he/she prefers. Based on $N$ left/right choices, a genetic algorithm produces a successive generation of $N$ vectors, in a constant feedback loop of offspring generation and {\it selection} operated by the subject. The iteration of this process leads to a sequence of $T$ generations of facial vectors, each one more adapted than the last to the subject's selection criteria, eventually converging to a pseudo-stationary regime in which the populations are similar to themselves and among consecutive generations. Fig.~\ref{fig:convergence} reports the evolution (versus the generation index, $t=1,\ldots,T=10$) of the {\it intra-population distance}, the distance among faces within the single populations sculpted by $10$ different, randomly chosen, subjects in E1 (see Supplementary section S4 for details). In the next subsection, we discuss the degree of reproducibility of our results as a function of $N$, $T$ and $S_1$.

The intra-population distance decreases with the generation index, indicating that the populations sculpted by single subjects tend to clusterize in a region of the face-space. This clustering is not observed in a null experiment in which the left-right decisions are taken randomly. Remarkably, a diversity of behaviors towards the pseudo-stationary regime is observed, already signaling differences in the way the face-space is explored. 

From now on, we will consider the final population sculpted by the $s-$th subject, $\{{\bf f}^{(s,n)}\}_{n=1}^{N}$, as the final, $T=10$-th generation of the sequence of populations sculpted by this subject in E1. In the next subsection we show that the face-space attractors of different subjects are actually significantly and consistently different. This experimental scheme is, therefore, able to {\it resolve} the subjective character of attractiveness, as the single subject tends to sculpt populations of vectors clustered in a narrow region in the face-space in successive realisations of the experiment. All these facts imply that the single subject attractor can be operationally characterised as an {\it extremum of a subject-dependent, probabilistic function in face-space}, which may be inferred from the populations sculpted by the subject in several instances of the experiment (see Supplementary Section S2 for a complete definition). The attractors are extrema of such a function in the sense that a significant fluctuation of a vector coordinate away from its value in the attractor will tend to lower its probability of being selected by the subject, given the reference portrait. 

\subsection*{Assessment of subjectivity: distinguishable aesthetic ideals}

In order to assess the subjectivity of the sculpting process, we need to measure to what extent the same subject, by repeating the same experiment, would sculpt populations of facial vectors closer to each other than to populations sculpted by distinct subjects. 
To this end we performed a second experiment (E2), in which a subset of $S_{\rm sc}=6$ subjects were asked to perform $m=6$ instances of an experiment E1, with the common reference portrait RP1, different (random) initial conditions and sequence of random numbers in the genetic algorithm. The subjectivity is assessed through the comparison of two sets of distances: (i) the ($S_{\rm sc}\, m(m-1)/2$) {\it self-consistency distances} among facial populations sculpted  by the same subject in different instances of the experiment E2; (ii) the ($S_1 (S_1-1)/2$) {\it inter-subject distances} between couples of populations sculpted by different subjects in experiment E1 (see Supplementary section S4 for details). If subjectivity was at play in the sculpting process, and not hindered by the stochasticity of the algorithm, the self-consistency distances would be lower than inter-subject distances.

\begin{figure}[t!]                        
\begin{center} 
	\includegraphics[width=.8\columnwidth]{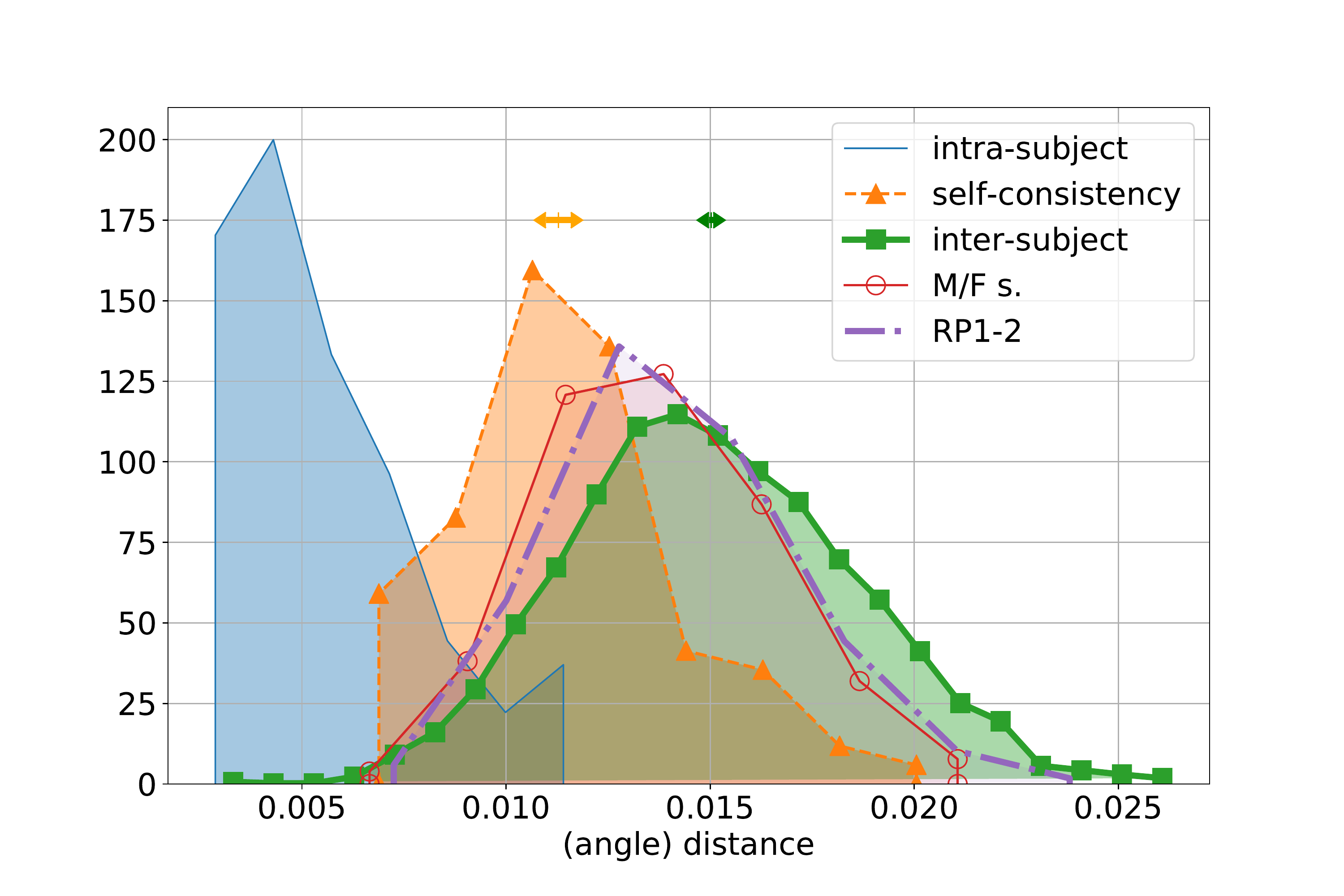}  
	\caption{Main panel: Normalised histograms of pseudo-distances.  {\bf Blue}: {\it subject intra-population distances}, or self-distances of all the populations sculpted in E1. {\bf Orange}: {\it self-consistency distances}, or distances among couples of populations sculpted by the same subject in E2. {\bf Green}: {\it inter-subject distances}, or distances among couples of populations sculpted by different subjects in E1.  {\bf Purple}: distances among couples of populations sculpted by different subjects in different experiments, E1 and E3 (differing in the reference portrait). {\bf Red}: distances among couples of populations sculpted by subjects of different gender in E1. The orange and green arrowed segments over the self-consistency and inter-subject histograms indicate the confidence intervals of the histogram averages, $\mu_{\rm sc}\pm\sigma_{\rm sc}/n_{\rm sc}^{1/2}$ and $\mu_{\rm i}\pm\sigma_{\rm i}/n_{\rm i}^{1/2}$ respectively, with $n_{\rm sc}=S_{\rm sc}m(m-1)/2$ and $n_{\rm i}=S_1(S_1-1)/2$. }
\label{fig:distances}
\end{center}   
\end{figure}

This is clearly the case, see Fig.~\ref{fig:distances}: self-consistency distances are lower than inter-subject distances (Student's $p< 10^{-30}$). In Fig.~\ref{fig:distances} we also report the histogram of {\it intra-population distances}, i.e., the average distance among the vectors belonging to a population, for different populations scuplted by different subjects in E1 (blue curve). The intra-population distances are not suitable for an assessment of the subject self-consistency, since they strongly depend on the number of generations performed by the genetic algorithm (c.f. Fig.~\ref{fig:convergence}). The emerging scenario is that of single subjects who, in a  single realization of the sculpting experiment, end up in a very clustered population (blue curve in Fig. ~\ref{fig:distances}). Performing several realizations of the same experiment leads the subject to a slightly different population in face-space (orange curve in Fig.~\ref{fig:distances}, labelled ``self-consistency''). These self-consistent populations are anyway closer to each other than to populations sculpted by different subjects, as witnessed by the larger {\it inter-subject distances}, whose histogram is presented in the green curve in Fig.~\ref{fig:distances}. A crucial point is that the distance between the inter-subject (green curve, ${\rm i}$) and self-consistency (orange curve, ${\rm sc}$) histograms in Fig.~\ref{fig:distances}, $t=(\mu_{\rm i}-\mu_{\rm sc})/(\sigma_{\rm i}^2+\sigma_{\rm sc}^2)^{1/2}=0.82(1)$ (see Supplementary Fig. S3) would be {\it even larger} in an experiment with a higher number of generations $T$. Using larger values of the genetic algorithm parameters $T$ and $N$ would result in a lower value of $\mu_{\rm sc}$, at the cost of a larger experimental time, since $NT$ binary choices are required from the subject (see Sec. Methods and Supplementary Sec. S3). Furthermore, larger values of $S_1,S_{\rm sc},m$ would give rise to a lower statistical error of the considered observables (see Supplementary Sec. S4), proportional to $1/\sqrt{S_{1,{\rm sc}}}$ and, in particular, to an even more significant difference among both histograms, since the uncertainty of their average is proportional to  $\sigma_{\rm i}/S_1$ and $\sigma_{\rm sc}/\sqrt{S_{\rm sc} m}$, respectively. In any case, the values used in experiments E1-2 are large enough to assess the differences among different subjects' attractors in a significant way.

The set of populations ${\cal S}_1=\{{\bf f}^{(s,n)}\}_{s,n}$ sculpted in E1 exhibits facial coordinates which vary in a wide range: roughly $0.018(10)$ per coordinate of the total face length, corresponding to $\sim 3.2{\rm  mm}$ in the average female face \cite{farkas1994} (see the average $\<{\bf f}\>$ and standard deviation $\bm \sigma$ of the single coordinates in Supplementary Fig. S5). The self-consistency distance $\mu_{\rm sc}\pm\sigma_{\rm sc}$, with which the experiment allows to resolve the single-individual attractor is, remarkably, much lower, equal to $0.0067(18)$ per coordinate (using the simple Euclidean-metrics in face-space, see Supplementary section S8), barely twice the pixel image resolution, $\sim 400^{-1}$ (figure Supplementary section S4). This quantity corresponds to  $1.18(30){\rm mm}$ in the female average facial length. 

Several metrics among facial vectors have been used to compute the inter-subject and self-consistency distances: {\it Euclidean, Mahalanobis, angle-} and {\it Byatt-Rhodes metrics} (see Supplementary section S4 and \cite{hill2011,valentine2016}). The angle-metrics (the angle subtended among standardised Principal Components (PC's) in face-space) turns out to be the one with which the statistical distinction is more significant (see Supplementary Fig. S3, and subsection ``Differences induced by the subject gender'' for the definition of PC's). This result is compatible with previous work proposing that such face-space metrics is the one that best captures differences in facial identity \cite{burton2001,hill2011}. Further results regarding the $t$-value difference among both histograms as a function of the face-space metrics can be found in the Supplementary section S4. Using the simple Euclidean metrics (the Euclidean distance per coordinate in physical coordinates), the inter-subject and self-consistency distances result slightly more overlapping, although still clearly distinct. For the sake of the statistical discernibility among the inter-subject and self-consistency distances, it is observed that the 10 dimensions involved in the definition of the face space are redundant in the sense that defining the face-space metrics in terms of the 7 most varying PC's, the two sets of distances result more significantly different (see Supplementary Fig. S3). 

For completeness, in Fig.~\ref{fig:distances} we also report two further sets of distances. The red line histogram corresponds to pseudo-distances among pairs of populations sculpted by subjects of different gender in E1, while the purple line histogram corresponds to the pseudo-distances among pairs of populations sculpted by different subjects with different reference portraits (E3, see "Relevance of facial features", before). 

These findings highlight the intrinsic subjectivity of facial attractiveness. Despite the limited freedom of choice, the reduced dimension of the face-space, and the common reference portrait, single subjects tend to sculpt a region of face-space that is systematically closer to their previous selections than to other subjects' sculptures. Indeed, the probability of two facial vectors sculpted by the same subject to be closer than two facial vectors sculpted by different subjects in E1 is $p_{12}=$\overlappingprobability$\,$  (see Supplementary section S8). 

A further interesting observation about Fig.~\ref{fig:distances} concerns the overlap between the histograms of {\it self-consistency} and {\it inter-subject distances}. Its existence allows us to reconcile the strong subjectivity unveiled by experiments E1-2, and the universality reported in the literature. The couples of facial vectors which are involved in distances for which there is a high overlap correspond to commonly preferred faces, around the most probable vector in the dataset, $\<\f\>$. Within a low experimental precision, or an accuracy larger than the standard deviation per coordinate $a>|{\bm \sigma}|/D$, all the subjects appear to agree in their choices. Under this perspective, the reported universality of beauty could be the side-effect of an experimental procedure where subjects express their preferences among a limited set of predefined options, the real facial images, in a high-dimensional face-space (indeed, the effective number of relevant facial dimensions may be of the order of hundreds \cite{chang2017}). In such an undersampling situation, different natural faces exhibit very different number of facial coordinates $g_i$ (or, more precisely, of PC's, see before), close to the most probable value $\<g_i\>$, with respect to their standard deviation (say, $\sigma(g_i)$). The faces exhibiting many coordinates in the commonly preferred region are consensually preferred, and most highly rated \cite{valentine2016}. By letting the subjects sculpt instead their preferred modification in a lower-dimensional face space, as in experiments E1-2, the subjects exclude extreme values of the coordinates, and manage to fine-tune them according to their personal criterion. In this circumstance, it is possible to resolve the subjects' preferences with higher accuracy, $\mu_{\rm sc}<|{\bm \sigma}|/D$, unveiling a strong subjectiveness. Our data suggest that the higher the accuracy with which the single subject attractor is resolved, the more distinguishable different subjects' attractors result in the face space. This picture suggests a {\it complete subjectivity}, or complete distinctiveness of different subjects' criteria (see also Sec. Methods).  

\subsection*{Correlations among different facial features \label{sec:correlations}}

In our experimental scheme, only geometric degrees of freedom may change. This allows us to determine the personal attractors efficiently and accurately, in a not too high-dimensional face-space. Moreover, it avoids the uncontrolled influence of features not described in the face-space. However, as anticipated in Sec. Introduction, it is also essential in this framework to account for possible mutual dependencies between different components of the facial vectors.

Besides the average and standard deviation of single coordinates referenced above, a quantity of crucial importance, despite the scarce interest that the literature has dedicated to it, is the correlation among facial coordinates from subject to subject. We denote with $\y$  the standardised fluctuations of the vector $\bf f$ around the experimental average, $y_i=(f_i-\< {f_i}\>)/\sigma_i$. The sculpted facial vectors presenting a fluctuation of a coordinate $y_i$ (say, a larger mouth width, $y_7>0$ in terms of inter-landmark distances) {\it tend to consequently present positive and negative fluctuations of other facial coordinates $y_{j\ne i}$} (e.g., a higher mouth, $y_4>0$). The sign and magnitude of such covariations is given by the {\it correlation matrix} among fluctuations of facial coordinates. This is the positive definite, symmetric matrix $C_{ij}=\< {y_i y_j}\>$, averaged over subjects $\<{\cdot}\>=\sum_{s=1}^{\Nv}\cdot/\Nv$. In order to subtract the influence of correlations within the single-subject attractor, only one population vector, of index $n_b(s)$, uncorrelated and randomly distributed, is considered for each subject $s$; the average and standard deviation of the matrix elements $C_{ij}$ have been obtained from many bootstrapping realisations, labelled by $b$, of the indices $n_b(s)$, see Supplementary section S4. The experimental matrix $C$  exhibits a proliferation of non-zero elements ($32\%$ of the matrix elements presenting a $p$-value $<5\cdot 10^{-2}$, see Supplementary section S11), unveiling the presence of strong correlations among several couples of facial coordinates. 

The most strongly correlated $C$ elements are among vertical or horizontal distances (see Supplementary Fig. S9 and table S4). Such strong correlations are easily interpretable: wider faces in ${\cal S}_1$ tend to exhibit larger inter-eye distances and wider mouths and jaws; higher nose endpoints, in their turn, covary with higher mouths and eyes; higher eyes covary with higher mouths, and so on. Perhaps the most remarkable aspect of the matrix $C$ is the proliferation of couples of vertical-horizontal coordinates,  highlighting the crucial role played by {\it oblique} correlations. The sign of oblique correlations $C_{ij}$ (see Supplementary table S4) is such that fluctuations of a landmark position $\vec\ell_\alpha$ covary with fluctuations of different landmarks $\vec\ell_\beta$ in such a way some inter $\alpha,\beta$-landmark segment slopes are restored  with respect to their average value. This is so for the most correlated couples of vertical-horizontal coordinates $i,j$ ($p<5\cdot 10^{-2}$).

\begin{figure}[t!]                        
\begin{center} 
\includegraphics[width=.35\columnwidth]{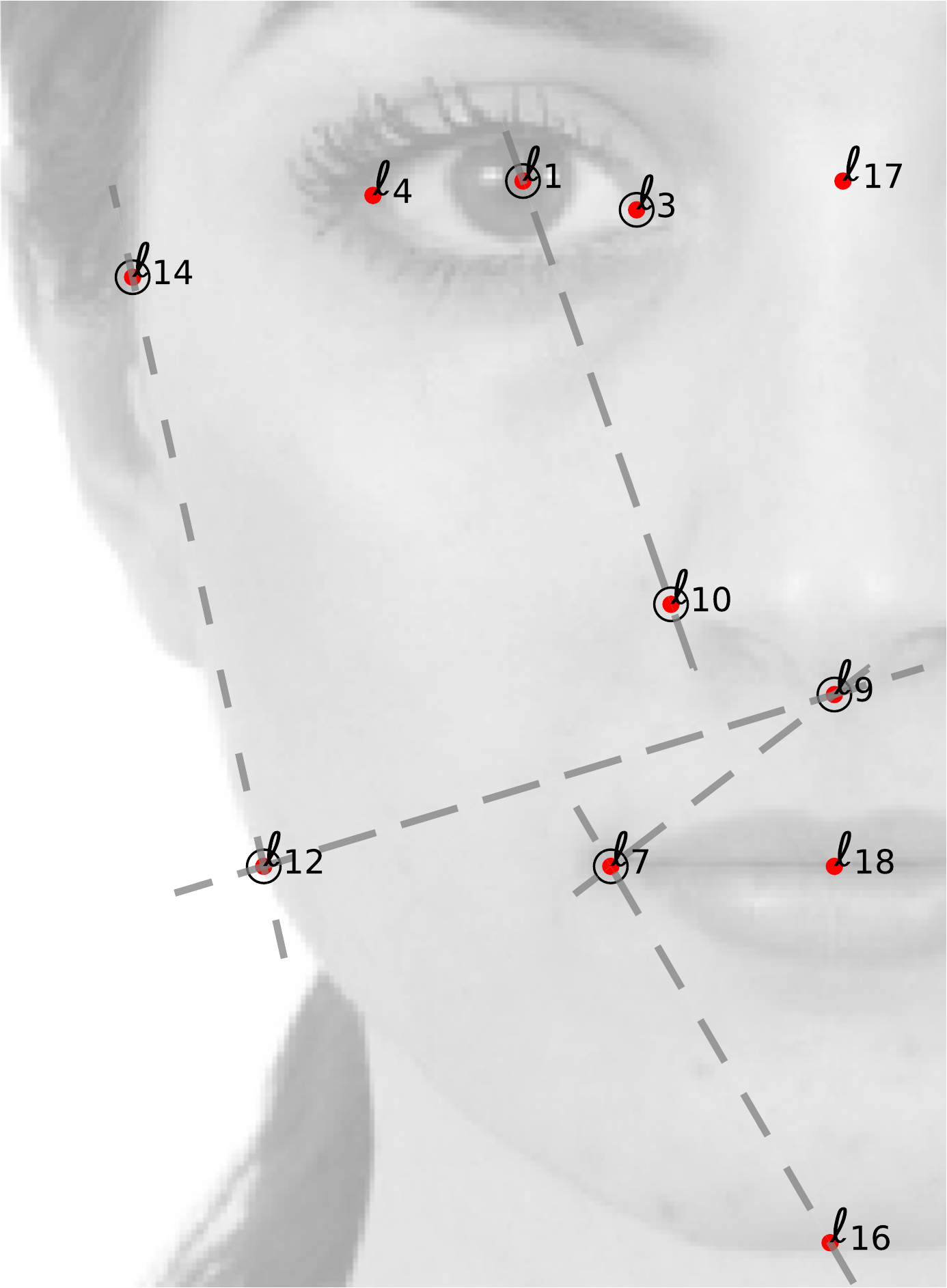}  
\caption{
	Relevant inter-landmark segments. The correlation matrix elements $C_{ij}$ involving vertical and horizontal landmark coordinates, $\<{\sf x}_{\alpha(i)}{\sf y}_{\alpha(j)}\>$ can be understood geometrically as a statistical invariance of the value of some inter-landmark segment slopes (dashed lines) with respect to their average value (represented in the figure). The sign of oblique $C_{ij}$'s coincide with that of the slope of the inter-landmark lines $(\<{\sf y}_{\alpha(i)}\>-\<{\sf y}_{\alpha(j)}\>)/(\<{\sf x}_{\alpha(i)}\>-\<{\sf x}_{\alpha(j)}\>)$. For instance, the most correlated horizontal-vertical landmarks are $\<{\sf x}_{12} {\sf y}_{9}\>$,  exhibiting a positive sign (c.f. Supplementary table S4): indeed, for lower nose endpoints (which correspond to a positive fluctuation  ${\sf y}_{9} > \<{\sf y}_{9}\>$), the $9-12$ angle can be restored only by increasing the ${\sf x}_{12}$-coordinate, ${\sf x}_{12} > \<{\sf x}_{12}\>$.
}
\label{fig:relevantangles}
\end{center}   
\end{figure}

The information brought by the correlation matrix helps in this way to construct a remarkably clear picture of the experimental distribution of facial vectors. The inter-subject differences and the experimental stocasticity induce fluctuations around the average facial vector $\y={\bf 0}$. The fluctuations are, however, strongly correlated in the facial coordinates, in such a way that vertical and horizontal coordinates covary positively and, {\it at the same time}, the value of  some inter-landmark segment slopes shown in Fig. \ref{fig:relevantangles}, of prominent relative importance, do not change too much with respect to their average value (see  Supplementary section S13).

These findings indicate that, for a meaningful inference of the perceived attractiveness in face-space, one should consider the impact of at least {\it linear combinations} of facial coordinates, rather than the impact of single facial coordinates.  The intrinsic complexity of attractiveness perception cannot be satisfactorily inferred through a simple regression of facial datasets using a sum of functions of single facial coordinates (see also Supplementary section S14 and \cite{ibanezberganza2019}).

\subsection*{Relevance of facial features: the variable hierarchy \label{sec:robustness}}

In this section we discuss the robustness of the results presented above. One of the crucial questions in facial attractiveness is what is the relevant set of variables which mainly determine the perceived attractiveness of a face \cite{laurentini2014,adolphs2016}. A formulation of the problem in theoretical-information terms is that of finding a hierarchy of relevant facial features. It is such that, when enriching the description with more variables in high levels of the hierarchy, the resulting variables in lower levels result unchanged. In the present study, the geometric quantities can be considered as low-level variables in the extent to which they are not influenced by the reference portrait, or by the luminance and texture facial features that have been disregarded and kept unchanged in the face-space description.

To settle this question, we performed a third experiment, dubbed E3, in which we asked the  $S_1$ participants in E1 to repeat the experiment using a different reference portrait (RP2, see Fig. \ref{fig:facespace}-D). Afterwards, we have compared the resulting set of sculpted facial vectors, ${\cal S}_3$, with the outcome of experiment E1, ${\cal S}_1$. Interestingly, a statistical $t$-test shows that, while some facial coordinates result clearly distinguishable, others result statistically indistinguishable, signifying their robustness with respect to the texture facial features determined by the reference portrait. These are, in terms of inter-landmark distances, $d_i$, the coordinates $d_{2,6,7,10}$, indistinguishable with $p>0.1$ (see Supplementary Fig. S6). If, instead of focusing on the distribution of single quantities $y_i$, one considers instead the correlations, $y_i y_j$, the results (see Supplementary table S4) turn out to be robust within their statistical errors, since only $2\%$ of the matrix elements $C_{ij}$ result significantly distinguishable ($p<0.075$, and none of them for $p<0.05$).

The  ensemble of these results implies a strong robustness of the results presented above, namely the subjectivity and the correlations among different facial features, with respect to a change in the reference portrait. It is remarkable that the coordinates $i=2,6,7,10$ in ${\cal S}_1$ are indistinguishable from those in ${\cal S}_3$ up to a remarkably small scale. For them, the average difference of couples of coordinates, $\<f^{(s)}_i-f_i^{(s')}\>_{s,s'}$  (with subjects $s,s'$ belonging to E1 and E3, respectively), vanishes up to small fluctuations, lower than the statistical error of such quantity. Such an error, of order  $(S_1 S_3)^{-1/2}$, see Supplementary section S10, is: $\sigma({\<f_i^{(s)}-f_i^{(s')}\>})=1.54\cdot 10^{-2}$ per coordinate, which corresponds to $0.27{\rm mm}$ in the average female face. We consider this result as one of the most remarkable of the present work. It highlights the striking robustness of the inter-landmark distances $d_{2,6,7,10}$. Such variables are, therefore, in low levels of the variable hierarchy, suggesting that they have prominent and intrinsic importance in the cognitive mechanism of face perception. 

\subsection*{Differences induced by the subject gender \label{sec:MF}}

An extensively debated question in the literature is to what extent the subject gender influences attractiveness, a question that the present experimental scheme is particularity suited to address. Partitioning the dataset  accordingly, ${\cal S}_1={\cal S}_{\rm m}\cup {\cal S}_{\rm f}$, it is  obtained that, again, some facial coordinates are barely distinguishable or completely indistinguishable in both sets ($d_{3,4,6,7}$, see Supplementary Fig. S7). Conversely, some coordinates are noticeably distinguishable. Compared to female subjects, male subjects tend to prefer thinner faces and jaws ($d_{5,10}$), lower eyes ($d_1$), higher zygomatic bones ($d_{0}$),  larger eye width ($d_8$). The difference becomes very distinguishable along  $d_{2,9}$ ($p<3\cdot 10^{-3}$, Supplementary Fig. S7): males definitely prefer shorter and thinner noses. These results are partially in agreement with previous findings in the literature, that highlight male subjects' preference for smaller lower face area and higher cheekbones \cite{perrett1994,rhodes2006}. Furthermore, they also provide accurate relative differences along each coordinate and reveal that, at least for the two reference portraits RP1-2, the facial feature leading to larger differences among men and women attractors {\it is the nose}.

A deeper insight is obtained by the analysis of PC's. These are the projections of the physical coordinates on the $C$-matrix  eigenvectors, ${\bf y}'=E{\bf y}$ (where $ECE^\dag={\rm diag}(\lambda_1,\ldots,\lambda_D)$). The different principal components $y'_i$ are, in other words, uncorrelated linear combinations of the physical coordinates ($\<y'_iy'_j\>=\lambda_i\delta_{ij}$). Principal components corresponding to large eigenvalues (as $y'_{10}$) represent the linear combinations of physical coordinates accounting for as much of the database variability, while those corresponding to the lowest eigenvalues represent the most improbable, or ``forbidden'' linear combinations of fluctuations away from the average ${\bf y}={\bf 0}$ (see the Supplementary Information). Different principal axes (${\bf e}^{(k)}$, the rows of matrix $E$) describe the different, independent sources of variability in the dataset, that could reflect the subjects' traits  most distinguishing their aesthetic criteria (as the gender).

\begin{figure}[t!]                        
\begin{center} 
\includegraphics[width=.8\columnwidth]{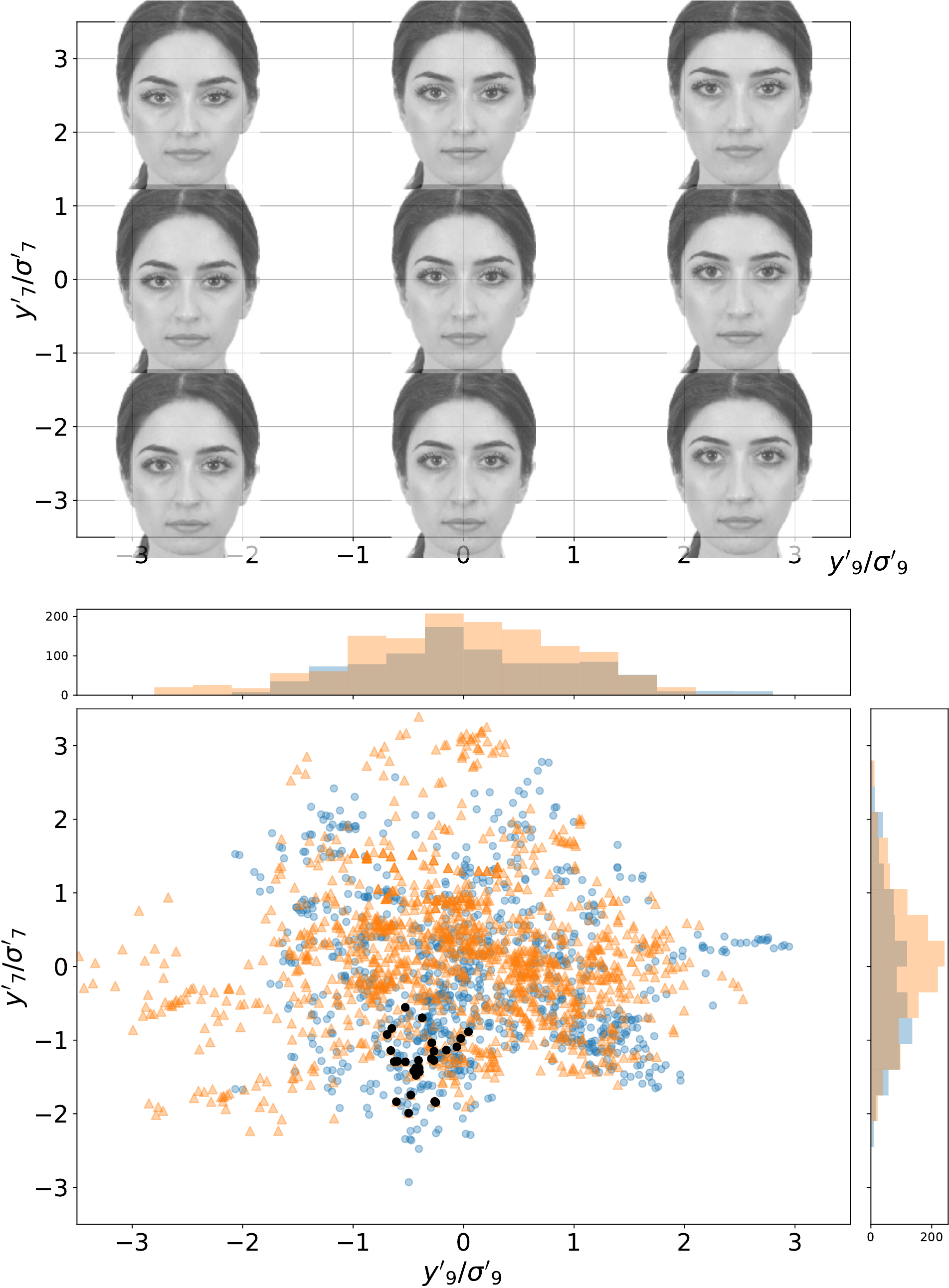}  
\caption{
	Top figure: facial images corresponding to the deformation of the average facial vector along two different principal axes (the ${\bf e}^{(7)}$, ${\bf e}^{(9)}$ eigenvectors of the correlation matrix $C$, corresponding to the fourth  and second larger eigenvalues, $\lambda_7$, $\lambda_9$). The axes represent the principal components along these axes, ($y'_9$, $y'_7$) in units of their standard deviations ($\lambda_i^{1/2}$). In other words, the image is generated from the facial vector ${\bf y}=E^\dag(y'_7 {\bf e}^{(7)}+y'_9 {\bf e}^{(9)})$.  Bottom figure: selected facial vectors. Each point is a projection of a selected facial vector in the principal axes corresponding to the Top figure, i.e., each point has coordinates ${y'}_7^{(s,n)},{y'}_9^{(s,n)}$, for all $s,n$ in the  E1 dataset. Blue points correspond to male subjects, and orange triangles to female subjects (male subjects tend to sculpt vectors with $y'_9<0$, and vice-versa). The black points correspond to a population sculpted by a single, randomly selected, subject. 
}
\label{fig:eigenvectors2D}
\end{center}   
\end{figure}

It turns out that faces corresponding to different subject's gender are distinguishable on three PC's (see Supplementary Fig. S8). Quite interestingly, such principal axes are not the ones exhibiting the largest eigenvalue, suggesting that the largest differences among selected faces correspond to inter-subject {\it criteria} that are transversal with respect to the subject's gender. Fig.~\ref{fig:eigenvectors2D} shows some image deformations of the average face along two principal axes: ${\bf e}^{(9)}$, ${\bf e}^{(7)}$ (the 2nd and the 3rd most variant eigenvectors of $C$). The PC defined by ${\bf e}^{(9)}$ is male/female distinguishable (males preferring negative values of $y'_9$ ). Instead, the $y'_7$ coordinate is gender-indistinguishable, and it could correspond to a different subject's quality, as the predilection for assertiveness, neoteny, or a different personality dimension, in the language of the {\it multiple motive hypothesis} \cite{cunningham1995,edler2001,little2014}.

\section*{Discussion \label{sec:discussion}}

In this article, we have introduced an experimental behavioural method that allows human subjects to efficiently select their preferred modification of a reference portrait in the multi-dimensional face-space (and, in principle, in general spaces of images that can be parametrised with 2D landmark coordinates). The method allows to flexibly and accurately determine the face-space regions which are representative of a given subject's criterion. It opens the path to a novel, data-driven approach to cognitive research in face perception, allowing scholars to: (1) quantitatively address the inter-subject differences {\it in the resulting sculpted shapes, beyond the rating}; (2) isolate the influence of a {\it secondary} set of variables (such as texture features) and {\it a posteriori} address their influence (something that cannot be directly done with databases of natural facial images); (3) analyse a resulting set of facial vectors without being limited or conditioned by the {\it a priori} correlations present in natural image databases.

The method ({based on our software FACEXPLORE, whose details are explained in the Supplementary Information}) permits a highly accurate description of the single subject or subject category preferences in the face-space, thanks to the geometric/texture separation of facial degrees of freedom and to a genetic algorithm for efficient search in the face space. Using this technique, we have performed a set of experiments in which the single subjects preferred region in the face space have been determined with an unprecedented accuracy, below the millimeter per facial coordinate.  

Such experiments allow us to draw the following conclusions. First of all, attractiveness turns out to be associated with the existence of subject-dependent specific regions in the face space that we dubbed attractors, highlighting the essential subjectivity of attractiveness. Despite the limited face-space dimension, and the homogeneity of the statistical universe (composed of subjects of the same cultural group), different subjects clearly tend to prefer different facial variations, suggesting that the subjectivity should be taken into account for a complete scientific picture of the phenomenon. Larger databases and more heterogeneous statistical universes would only make the essential subjectivity of attractiveness perception even more evident.

In light of these facts, the validity of the natural selection hypothesis (universality, impact of averageness, symmetry and sexually dimorphic traits) may be arguably a matter of the precision of the length scale and of the facial image resolution of the facial description. Within a sufficiently accurate description of the subjects' criterion in face-space, the phenomenon emerges in its whole complexity, showing that the preferred faces of different subjects are systematically different among themselves and, consequently, different from the average face. In their turn, these differences reflect personal features and circumstances that condition the subject’s preferences, one of which is the subject's gender.

The second important conclusion we can draw concerns the patterns associated to different subjects' attractors. Different sculpted facial vectors exhibit strong correlations among pairs of facial distances, characterising the underlying universality and complexity of the cognitive processes, leading, in its turn, to the observed subjectivity \cite{little2014}. Our study reveals, in particular, the crucial importance of correlations among vertical and horizontal coordinates, whose existence and relevance have been, to the best of our knowledge, only postulated \cite{pallett2010, joy2006, galantucci2014}. Different facial variations are strongly 
correlated, a fact that confirms the holistic way in which we perceive faces (see references in \cite{adolphs2016}). Our results suggest to consider attractiveness not as a scalar quantity, rather as the outcome of a complex process in which various semantic motives are evaluated. These are probably encoded in pairwise and higher-order correlations among facial features, more than in the value of single facial coordinates \cite{galantucci2014}.

A third result concerns the role of the subject's gender in the assessment of attractiveness. This is, indeed, an important source of diversity in our dataset. Nose length and width, eye height, face and jawbone width, zygomatic bone height, turn out to be the main facial traits distinguishing male and female observers. However, a principal component analysis suggests that the largest differences among selected facial variants correspond to principal axes that are independent of the subject's gender. Abstract personality dimensions have been observed to be consensually attributed to faces, and the impact of such qualities on various facial elements have been measured through principal component analysis \cite{oosterhof2008, todorov2011, walker2016, abir2017}. Such principal axes could be correlated with those of the present study. This would be a confirmation of the postulated connection between attractiveness and personality judgments \cite{bzdok2011, walker2016, little2006}. It would allow to elicit the different traits that are judged by the subjects in a bottom-up, data-driven fashion.

A further noticeable result is the assessment of the influence of the reference portrait in the distribution of sculpted facial vectors. Quite remarkably, the {\it a priori} dimensionality reduction implicit in our analysis (ignoring texture degrees of freedom), turns out a posteriori to be sufficient and justified (see Sec. Methods).

In summary, the novel experimental approach proposed in this article allowed us to unveil the essential subjectivity of attractiveness. The subjectivity emerges more evidently in the present scheme, since the reduction of the number of face space dimensions allows to avoid the undersampling occurring in experiments in which the subjects are asked to choose or rate natural faces. 

We believe that the generality and reliability of the present approach could have a strong impact on future studies about beauty and pleasantness in different domains.

Possible completions of the present work are: an assessment of the robustness of principal components; an analysis of the intra-subject correlation matrix of facial coordinates; a variant of the analysis of correlations in an experiment with real facial images (whose landmarks could be automatically identified with deep learning techniques \cite{wang2017}); an unsupervised inference analysis of the database (already being carried on in our group) within the framework of the Maximum Entropy method.

\section*{Methods \label{sec:methods}}

\def\f{\textbf{f}}

\subsection*{Face space}

Our experimental design is based on the parametrisation of the face in a $10$-dimensional face-space defined by $D=11$ vertical and horizontal {\it inter-landmark distances}, ${\bf d}=(d_i)_{i=0}^{D}$ between standard facial landmarks (see figure \ref{fig:facespace}-A). The inter-landmark distances are subject to a constraint ${\sf h}=\sum_{i=1}^4 d_i=1$, reflecting the intrinsic scale invariance of the problem, in such a way that all distances $d_i$ are in units of the total facial length (i.e., they represent {\it proportions with respect to the facial length}, rather than absolute distances). 
 As vector of facial coordinates $\f$, we have considered both the $11$ distances $f_i=d_i$ themselves or, alternatively, the non-redundant (and unconstrained) subset of $D=10$ {\it Cartesian landmark coordinates} of a set of landmarks $\vec\ell_\alpha=({\sf x}_\alpha,{\sf y}_\alpha)$ (with $\alpha=1,3,7,9,10,12,14$, see figure \ref{fig:relevantangles} and Supplementary Sec. S13), that can be unambiguously retrieved from the set of inter-landmark distances. All the results presented in the article are qualitatively identical using the inter-landmark distances $d_i$ or the landmark Cartesian coordinates $\vec\ell_\alpha$ as facial vectors.

\subsection*{Separation of geometric and texture degrees of freedom\label{sec:separation}}

The face-space parametrisation is based, as previously mentioned, on the decoupling of {\it texture} (lightness, detailed, and skin textural) facial features, on the one hand, and {\it geometric} (landmark coordinates), on the other hand. The separation of these two kinds of degrees of freedom is a standard paradigm of face representation (see, for example, \cite{abir2017,chang2017,walker2016}). It has been argued, in the light of the recently decoded neural coding for the facial identity in the primate brain, to be a naturally efficient parametrisation of the face \cite{chang2017}, outperforming other techniques in which texture and landmark-based are not separated, as the   description in terms of {\it eigenfaces}.

\subsection*{Image deformation \label{sec:imagedeformation}}

Given a reference portrait (see figure \ref{fig:facespace}-B) and a vector of facial distances ${\bf d}_1$, we create, by means of  image (similarity transformation) deformation algorithms \cite{schaefer2006}, a realistic facial image based on the reference portrait, deformed in such a way that the inter-landmark distances defined in figure \ref{fig:facespace}-A assume the desired values, ${\bf d}={\bf d}_1$. Given the reference portrait image ${\cal I}_0$, the position of its corresponding landmarks $\vec\ell_{0,\alpha}$, and the vector $\bf d$, we calculate  the Cartesian coordinates $\vec\ell_{1,\alpha}$ of the new set of landmarks, completely defined by $\bf d$. The image deformation algorithm then generates a new facial image  ${\cal I}_1$  with a point-dependent parameter linear transformation, such that the pixels occupying the landmark positions $\vec\ell_{0,\alpha}$ in the original image are mapped into the new positions $\vec\ell_{1,\alpha}$, and the rest of the pixels of the original image are mapped in order to produce a resulting image as realistic as possible. We have observed that, in order to produce realistic results, the linear transformation should be in the {\it similarity class} \cite{schaefer2006}, beyond {\it affine} transformations. The deformed image is actually not created by mapping every pixel of the original image, but only the corners of a sub-grid; the sub-images inside each sub-grid are then warped to a polygon defined by the mapped corners of the grid, through affine transformations. The size of the sub-grid is taken to $\lesssim 15$ pixels. Both the reference portrait and the deformed images are roughly $300\times400$ pixels for RP1-2.

\subsection*{Genetic algorithm of face-space exploration\label{sec:facexplore}}

The genetic algorithm  is based on a sequence of pairwise subject's choices among two facial images that are adaptively proposed to the subject, learned from his/her previous choices. An initial {\it population} of  $N$  vectors of randomised facial coordinates, ${\bf f}^{(s,n)}(0)$, evolve by means of genetic mutation and recombination, subject to the {\it selection} exerted by the experimental volunteer. At the $t$-th generation, the $N$ vectors of the population generate an offspring of $N$ individuals, by mutation and recombination according to the {\it differential evolution algorithm} (see Supplementary Sec. S3). The offspring is generated from the facial vectors only, independently of the reference portrait. The subject plays then the role of the evolutive pressure in the algorithm dynamics, selecting ($N$ times) one among {\it two facial images}: one made from a vector of the population (and a reference portrait), and one made from its offspring. The $t+1$-th generation of vectors is then taken as the $N$ vectors selected by the subject at the $t$-th generation. After a certain number, $T$, of generations, the population of facial vectors eventually reaches a regime in which the population of vectors do not change too much from one generation to the next. The $T$-th population of facial vectors is taken as the population of vectors sculpted by the subject, and constitutes the outcome of experiments E1-3.

This approach differs from previous approaches to facial attractiveness based on genetic algorithms \cite{johnston1993,wong2008} in what: it allows a subject to select {\it in real time} a {\it realistic} facial image; in terms of geometric quantities only; with {\it fixed texture degrees of freedom}; 
finally, {\it avoiding the use of numerical ratings}, since the subject performs a sequence of left/right choices rather than assigning ratings to the images.

Populations of facial vectors sculpted by different subjects tend to be more far apart than populations sculpted by the same subject (see Sec. Results). Remarkably, the real difference between different subjects' attractors is even larger, since it is unavoidably underestimated in virtue of the finiteness of the experimental method. Indeed, two standard deviations with different origins contribute to the self-consistency distance $\mu_{\rm sc}$ (see figure \ref{fig:distances}). One is the intrinsic, cognitive ambiguity of the subject's criterion; the other is the uncertainty brought by the genetic algorithm stochasticity (sec. Supplementary Sec. S3), whose origin is the discreteness of the proposed mutations and the consequent stochastic bias in the face space exploration. In genetic experiments with parameters in what we call in the {\it slow search regime} (mainly larger $N$ and number of generations, $T$), the algorithmic uncertainty decreases, and $\mu_{\rm sc}$ is expected to decrease consequently. This is the general expected behaviour of the differential evolution algorithm. We have also verified this fact experimentally: the distances among populations sculpted by a single subject significantly decrease for increasing values of $N=10,20,28$. As a consequence, variants of the present experiment with {\it slower} genetic algorithm parameters would more finely {\it resolve} different subject's facial ideals, leading to a larger gap between inter-subject and self-consistency distances,  at the cost of a larger number of subject's choices and experimental time. 

\subsection*{Details of the experiments}

Experiments E1, E2, E3 were performed by a pool of $\Nv=\nbsubjects$ volunteers ($54$ female, $39$ male, of age average and standard deviation: $26(12)$), mainly students, researchers and professors of the University ``La Sapienza''. Experiment E2 was performed under identical conditions of E1. A subset of $S_{\rm sc}=6$ participants to E1 (3 females, 3 males, of age average and standard deviation: $33(15)$), were asked to perform 5 further instances of the experiment E1, in five different days, using, as in E1, the reference portrait RP1. The genetic algorithm parameters used are (see Supplementary Sec. S3): $N=28$, $T=10$, $\mu=0.15$, $\rho=1$. Each subject performed a number of $NT=280$ choices among couples of facial images. These are  $400\times 300$ pixel, B/W images in an $1024 \times 768$ resolution monitor. The reference portraits RP1-2 have been taken from the {\it Chicago face database} \cite{ma2015}. Each experiment lasted roughly 25 minutes on average (see the histogram of time intervals among successive left-right choices in figure Supplementary Sec. S7). The subjects were asked to look away and relax for some second each $N=28$ choices. All methods in experiments E1-3 were carried out in accordance with relevant guidelines and regulations. The experimental protocols used have been approved by the General Data Protection Regulation (EU) 2016/679. Informed consent was obtained from all subjects. No subjects under 18 participated in the experiment.

\section*{Data Availability}
The data and the codes devoted to the data analysis are available by request to the corresponding author.

\section*{Acknowledgements}

We acknowledge  Andrea Giansanti, Davide Iannuzzi and Giovanni Pezzulo for suggestions and bibliographic highlights, Scott D. Shaeffer and William Schueller for technical and informatic advices, Andrea Gabrielli for mathematical suggestions, and Fernanda Pereira da Cruz Benetti for her help with the manuscript.

\section*{Author contributions}

M. I.-B. and V. L. devised the project. M. I.-B. devised the inference scheme and developed the FACEXPLORE software of face exploration and the data analysis software. M. I.-B- and V. L. organised the experiments. A. A. and M. I.-B. performed the data analysis. M. I.-B. and V. L. wrote the article.  

\section*{Additional information}
\textbf{Competing interests.} The authors declare no competing interests.

\bibliography{faces}



\end{document}


\date{}
\maketitle

\tableofcontents

\def\Ch{{C_{\rm h}}}
\def\Co{{C_{\rm obs}}}
\def\Cn{{C_{\rm null}}}
\def\Jh{{J_{\rm h}}}
\def\Jo{{J_{\rm obs}}}
\def\Jn{{J_{\rm null}}}

\section{Detailed description of the face-space\label{sec:facespace}}

\begin{figure}[t!]                        
\begin{center} 
\includegraphics[width=.4\columnwidth]{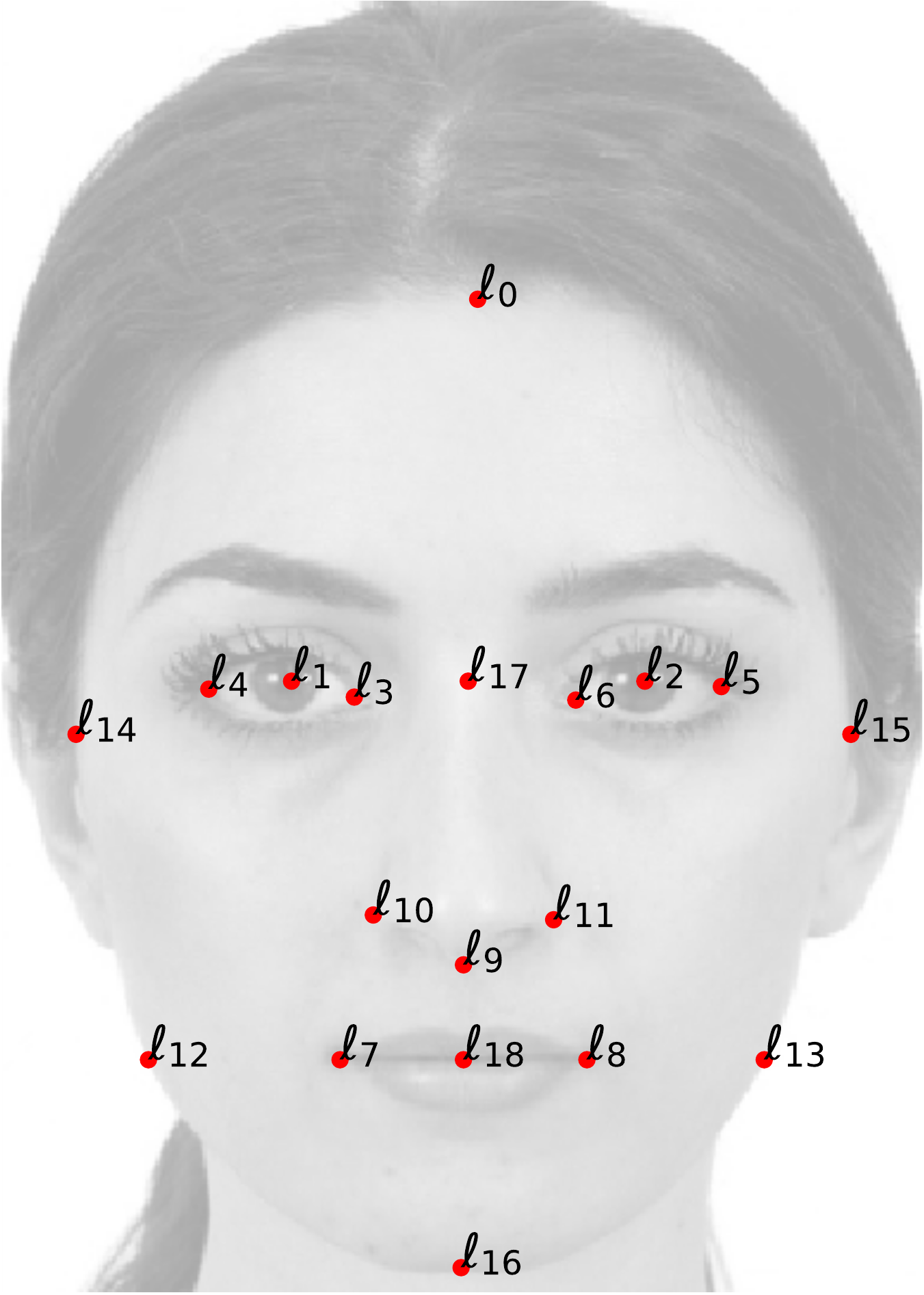}  
	\caption{Facial landmarks used to define the face-space. The numbers correspond to the index $\alpha$ of various  landmarks with Cartesian coordinates $\vec\ell_\alpha$.}
\label{fig:keylandmarks}
\end{center}   
\end{figure}

As explained in sections Results and Methods (main article), the face is parametrised as a vector ${\bf d}=(d_i)_{i=0}^{D}$ of $11$ inter-landmark distances, shown in figure 1-A (main article). These are the vectors defining the ``genoma'' of the population members in the genetic algorithm. In the {\it selection} step of the genetic algorithm (see section \ref{sec:geneticalgorithm}), a facial image is generated from every vector $\bf d$ in the genetic population. This is done through the generation of an auxiliary 36-dimensional vector of Cartesian landmark coordinates, ${\bf L}=(\vec\ell_\alpha)_{\alpha=0}^{18}$ (with $\vec\ell_\alpha=(\sx_\alpha,\sy_\alpha)$), obtained from $\bf d$. The vector $\bf L$ is a list of the ${\sf x},{\sf y}$ coordinates (in pixels, the $\sf y$ growing downwards in the image) of the various landmarks evidenced in figure \ref{fig:keylandmarks}, labelled by $\alpha$. The facial image corresponding to $\bf d$ is then generated, using the technique described in sec. Methods (main article), from the triplet $\{{\cal I}_0,{\bf L}_0,{\bf L}\}$, where ${\cal I}_0$ is the reference portrait image, ${\bf L}_0$ its Cartesian landmark coordinates, and ${\bf L}$ is the vector of Cartesian landmarks generated from the desired vector  $\bf d$. The mapping ${\bf L}\leftrightarrow{\bf d}$ is one-to-one, given the set of Cartesian landmark coordinates of the reference portrait being used, ${\bf L}_0$. The information present in ${\bf L}_0$ is used to impose some constraints: the eye aspect ratio (the $\vec\ell_1$--$\vec\ell_3$ segment slope $({\sf y}_3-{\sf y}_1)/({\sf x}_3-{\sf x}_1)$) is constant and equal to that of ${\bf L}_0$, and the same is valid for $\vec\ell_1$--$\vec\ell_4$ (otherwise the pupil could become an ellipse); ${\sf y}_9-{\sf y}_{10}$ is constant and equal to its value in ${\bf L}_0$. Moreover, the reference portrait determines the origin of the reference frame, the coordinate $\vec\ell_0$, which is fixed as well as $\vec\ell_{16}$. The information in $\bf L$ is highly redundant: by construction, the $\sf x$ coordinates of left/right landmarks are symmetric with respect to ${\sf x}_0$ and share the $\sf y$ coordinate; the 12-th landmark is defined in such a way that its ordinate coincide with that of $7,18,8,13$, and, analogously, ${\sf y}_1={\sf y}_{17}={\sf y}_2$. 

Let us now describe in detail the two parametrisations used to construct the face-space vectors $\bf f$. 

\begin{itemize}
\item {\it Inter-landmark distances.} The facial vector components  are taken as the  $11$ distances in $\bf d$

\begin{equation}
f_i=d_i
\end{equation}

 The names of the $11$ resulting facial coordinates are shown in table \ref{tab:interlandmarkdistances}. 
\item {\it Reduced set of Cartesian landmark coordinates.} The facial vector components are taken as a  vector of non-redundant, {\it reduced} set of $D=10$ Cartesian landmark coordinates, or $D$ non-redundant components of the vector $\bf L$. In other words:

\begin{equation}
	f_i={{\sf c}(i)}_{\alpha(i)}
,
\end{equation}
with $\alpha(i)=1,3,7,9,10,12,14$ and ${\sf c}(i)\in \{{\sf x},{\sf y}\}$, specified in table \ref{tab:cartesianlandmarkcoordinates}.
\end{itemize}

\begin{table}
\begin{tabular}[h]{cllll} 
$i$	&	type (h/v)  & definition & name \\
\hline
	0	&v	&	${\sf y}_{14}-{\sf y}_{0}$	& zygomatic bone ordinate \\
	1	&v	&	${\sf y}_{17}-{\sf y}_{0}$	& eye-forehead distance \\
	2	&v	&	${\sf y}_{9} -{\sf y}_{17}$	& nose length \\
	3	&v	&	${\sf y}_{18}-{\sf y}_{9}$	& nose-mouth distance \\
	4	&v	&	${\sf y}_{16}-{\sf y}_{18}$	& chin-mouth distance \\
	5	&h	&	${\sf x}_{15}-{\sf x}_{14}$	& face width \\
	6	&h	&	${\sf x}_{2}- {\sf x}_{1}$		& inter-eye distance \\
	7	&h	&	${\sf x}_{8}- {\sf x}_{7}$	& mouth width \\
	8	&h	&	${\sf x}_{3}- {\sf x}_{4}$	& eye width \\
	9	&h	&	${\sf x}_{11}-{\sf x}_{10}$	& nose width \\
	10	&h	&	${\sf x}_{13}-{\sf x}_{12}$	& jaw width 
\end{tabular}
\caption{Definition of the {\it inter-landmark distance} facial coordinates $f_i$ in terms of the horizontal/vertical (h/v) distances among landmark coordinates $\vec\ell_\alpha$ (see figure \ref{fig:keylandmarks}). \label{tab:interlandmarkdistances}}
\end{table}

\begin{table}
\begin{tabular}[h]{cllll} 
	$i$	&	$\alpha(i)$ &	${\sf c}(i)={\sf x},{\sf y}$  & name \\
\hline
0	&1	&{\sf x}& left pupil abscissa\\
1	&3	&{\sf x}& internal left eye limit abscissa\\
2	&7	&{\sf x}& left mouth limit abscissa\\
3	&10	&{\sf x}& left outermost nose limit abscissa\\
4	&14	&{\sf x}& zygomatic bone abscissa\\
5	&12	&{\sf x}& left jaw limit (at the mouth's height) abscissa\\
6	&1	&{\sf y}& left pupil ordinate\\
7	&7	&{\sf y}& left mouth ordinate \\
8	&9	&{\sf y}&  nose endpoint ordinate\\
9	&14	&{\sf y}& zygomatic bone ordinate 
\end{tabular}
	\caption{Definition of the {\it Cartesian landmark} facial coordinates $f_i={{\sf c}(i)}_{\alpha(i)}$ in terms of the landmark coordinates $\vec\ell_\alpha=(\sx_\alpha,\sy_\alpha)$ (see figure \ref{fig:keylandmarks}). For each coordinate $i$ we specify $\alpha(i)$ and ${\sf c}(i)={\sf x}$ or $\sf y$. \label{tab:cartesianlandmarkcoordinates}}
\end{table}

Both sets of coordinates such defined are equally dimensional: the inter-landmark distances $\bf d$ have one more degree of freedom, but are subject to the constraint ${\sf h}=\sum_{i=1}^4 d_i=1$, so that the dimensionality of both sets is 10. As previously stated, the mapping among both vectors is one-to-one: they contain the same information, and are such that there is no redundant information in the reduced $\bf L$ (given the reference portrait vector). 

All the observables that we have calculated in the data analysis (see section \ref{sec:observables}) can be computed in terms of $\bf d$'s or in terms of reduced $\bf L$'s. All the results presented in this article are qualitatively identical using either face-space parametrisation. Some results are clearer in terms of $\bf L$'s, due to the presence of the constraint ${\sf h}=1$ to which the  $\bf d$ vectors are subject. This is the case of the correlation matrix, that we have decided to show in figure \ref{fig:C} in terms of reduced $\bf L$'s (see below). The correlation matrix in terms of inter-landmark distances will be analysed in-depth in \cite{ibanezberganza2019}.

\section{Beauty as extrema in a given coordinate space \label{sec:experimentalaim}}

As explained in the main text,  our experimental setup allows a subject to sculpt a population of facial vectors, considered as an empirical sample of his/her attractor, or preferred region in face-space. The efficient characterisation of the attractor from a finite set of binary choices by the subject is an {\it inference} problem, that we tackle as an optimization problem, solved with the use of a genetic algorithm. This strategy is justified and motivated through the following assumptions. Given an experimental subject and a parametrisation of the human face in a real-valued vector, we assume that there exists a region of the face-space that represents the subject's preference, in the sense that he/she would statistically tend to prefer images in that region rather than those farther from it, and that this region can be probabilistically characterised within some accuracy. We postulate the existence of a subject-dependent probability distribution in the face-space, ${\cal L}_{\rm g}$, such that ${\cal L}_{\rm g}(\f)/{\cal L}_{\rm g}({\f}')$ represents the relative probability of the subject to express his/her preference for the facial image whose coordinates are $\f$ (so that a flat function represents a completely indifferent or unpredictable subject)
[{A convex function, locally flat around a set of points $\f^*$, $\nabla {\cal L}|_{\f^*}={\bf 0}$, $H=\det {\rm Hess} [{\cal L}](\f^*)<0$ would represent a subject which tends to refuse local variations away from such set, with a probability depending on the modulus of $H$
}.] If the subject could modify the coordinates of an image following his/her personal taste, he/she would tend to finally choose the face-space region corresponding to the relative extrema of the function ${\cal L}_{\rm g}$ (if  the facial image details, not parametrised by $\f$, are unchanged and given by the reference portrait).
Of course, in real experiments, the function  ${\cal L}_{\rm g}$ can only be inferred with uncertainty (induced, at least, by the subject's uncertainty), leading to an inferred function, $\cal L$. The present experimental scheme provides a finite set of representative vectors sampled with experimental uncertainty from ${\cal L}_{\rm g}$; hence, a function $\cal L$ may be inferred from such representative vectors, using the Maximum Entropy method \cite{ibanezberganza2019}. The inference quality depends on both the extent to which the set of populations is representative of the subject's attractor (a first-step, experimental inference), and on the inference procedure of the set of populations (the second-step inference, leading to $\cal L$). 
Under these working hypotheses, one is allowed to treat the inference of the single-individual attractor as an optimisation problem of a multi-valued function that, however, cannot be evaluated numerically. We now explain how this is done in our experimental scheme.

\section{Genetic Algorithm details \label{sec:geneticalgorithm}}

Let us specify the details of the genetic algorithm for face-space exploration. Each instance of the experiment is defined by a set of algorithm parameters, ${\cal P}=\{N,T,\rho,\mu\}$. 
The algorithm defines a stochastic, discrete time dynamics (the time index $t$) of the population of facial vectors $\f^{(s,n)}(t)$, $n=1,\ldots,N$, coupled to the dynamics of an abstract subject $s$, which performs binary choices among couples of vectors according to some stochastic rule. Even if the subject's binary choices were deterministic, the dynamics is intrinsically stochastic in the  initial condition and in the sequence of random numbers. {\bf (1) Initialization}. At $t=0$, the initial vectors ${\bf f}^{(s,n)}(0)={\bm\xi}_{(s,n)}+{\bf f}_0$ are taken as $N$ facial vectors whose coordinates are random, uncorrelated, zero-averaged fluctuations around a given (common to all subjects) facial vector ${\bf f}_0$. {\bf (2) Recombination and mutation}. For each of the $N$ facial vectors in the population, $\f$, a {\it child}, or potential offspring vector $\v$, is generated, according to a rule specified in the next paragraph (based on recombination   and stochastic mutation of the existing vectors in the population). {\bf (3) Selection}. For each of the $N$ couples  of vectors   of the original and of the offspring population, $\f,{\bf v}$, a pair of facial images ${\cal I}(\f)$, ${\cal I}({\bf v})$ is generated (with the image deformation algorithms described in sec. Methods (main article). Afterwards, the subject chooses  among the two images (the one corresponding to a vector belonging to the current generation, and the one corresponding to {\it its child}), for each of the $N$ pairs. The $N$  chosen facial vectors (of which some are offspring and some are parents in the $t$-th generation) will form the successive generation; {\bf (4)}. One now goes recursively to {\bf (2)} and $t+=1$ until $t=T$. \\ 

{\it Differential Evolution Algorithm.} The rules in step {\bf (2)} are given by a particular genetic algorithm called Differential Evolution Algorithm \cite{storn1997,tasoulis2005}. It has been chosen  due to its suitability to find multiple extrema and to the fact that it does not require the numerical evaluation of the function to be maximised, say $\cal L$, but only the boolean inequality ${\cal L}[\f\,] < {\cal L}[{\bf v}]$. In our experiment, the evaluation of this inequality corresponds to the choice of the subject between two images ${\cal I}(\f)$, ${\cal I}({\bf v})$.  Given the population at time $t$, the {\it son}, ${\bf v}^{(j)}$, of the $j-$th vector of the population $\f^{(j)}(t)$, is generated from this vector and from two different {\it parents}, $\f^{(j_1)}$, $\f^{(j_2)}$, with $1\le j_1\ne j_2\ne j \le N$, randomly selected. The mutation and recombination steps are, $\forall i=1,\ldots,D$:

\begin{displaymath}
 v_{i}^{(j)}= \left\{ \begin{array}{ll}
 f^{(j)}_{i}(t) + \mu\,(f^{(j_1)}_{i}(t)-f^{(j_2)}_{i}(t))& \textrm{ with prob. } \rho \\
 f^{(j)}_i(t)& \textrm{ with prob. } 1-\rho \\
\end{array} \right.
\label{eq:DEA}
\end{displaymath}
%
The selection and generation update steps can be written as:

\begin{displaymath}
 \f^{(j)}(t+1)= \left\{ \begin{array}{ll}
	 \v^{(j)}& \textrm{if } {\cal L}[\f^{(j)}(t)] < {\cal L}[{\bf v}^{(j)}] \\
 \f^{(j)}(t)& \textrm{otherwise}
\end{array} \right.
\label{eq:inequality}
\end{displaymath}

Note that the algorithm acts on every $i$-th single coordinate independently. $\rho$ is called the {\it crossover probability} and  $\mu$ the {\it mutation factor}, quantifying the amount of stochastic mutation in the genetic evolution. In our experiment, the evaluation of the inequality (\ref{eq:inequality}) corresponds to the choice between the corresponding facial images, ${\cal I}[{{\bf v}^{(j)}}]$ and ${\cal I}[\f^{(j)}(t)]$, by the subject. He/she chooses among the images corresponding to the first parent's vector and its child; the selected vector survives and becomes part of the successive generation. 

The FACEXPLORE software operates, in this way, in a regime defined by four parameters, $\rho$,  $\mu$, $N$ and $T$ (and by other algorithm details, such as the way in which the initial population of vectors is initialised, the reference portrait, the constraints imposed to the facial vectors at each generation, or the size of the sub-grid to be warped in the image deformation algorithm,  see sec. Methods (main article)). Especially for large values of $\rho$, and for small values of $\mu$, $N$ and $T$, the results of single realisations of the experiment depend on the sequence of random numbers and on the random initial condition used in the particular realisation. This is a general characteristic of the genetic algorithm, arising also in the optimisation of deterministic functions, not a specific characteristic of the FACEXPLORE software. For lower values of $N$, the algorithm does not perform an exhaustive local search in the parameter space at each generation. The offspring generation is biased by its finiteness, a fact that conditions the experimental course and, consequently, the outcome. Depending on the parameters ${\cal P}$, the algorithm stochasticity could become large enough to hinder the differences among different subjects' choices. The parameters can also be such that the stochasticity is moderate, in the sense that they allow to {\it resolve} the single subject peculiarities, whose existence has been demonstrated in the main article. Thus, for different values of the parameters $\cal P$, we can define two main situations:

\begin{itemize}
	\item For large $\rho$ and small $\mu$, $N$ and $T$ ({\it fast-search regime}): the algorithm ``converges fast'' (in the sense of Fig. 2 (main article)), but the resulting final populations vary significantly when the subject repeats the experiment. 
	\item For sufficiently small $\rho$ and sufficiently large $\mu$, $T$ and, above all, $N$ (what we call the {\it slow-search regime}): the experiment requires more choices to ``converge'', but the resulting population will respond more to the subject's criteria and less to the randomness, i. e., to the particular realization of the experiment.
\end{itemize} 

Ideally, at the end of the process, the population of vectors reaches a pseudo-stationary regime which is stable against local fluctuations, i. e. that the user does not want to change. For experiments deep in the slow-search regime, the {\it average} differences between the outcomes of different realizations of a single-subject experiment, $\mu_{\rm sc}$, should no longer depend on the algorithm parameters ($\mu_{\rm sc}$ would not decrease using {\it slower} parameters ${\cal P}$). In this case, the only experimental uncertainty would be the subject's uncertainty. In practice, such a deep slow-search regime would require a large number of choices. The algorithm parameters must satisfy a compromise between the desired accuracy and the time required by the subject to perform the experiment. 

A crucial point is how the initial population of vectors is selected. If the standard deviations of $\bm\xi$ are large enough, and for sufficiently large $N$, the initial population covers a broad region of the face-space (actually, the first-generation facial images often result grotesque and misshapen). Together with a large value of $N$, $T$, $\mu$ and $1-\rho$, it is expected that, again, the initial condition ${\bm\xi}_{(s,n)}$ and ${\bf f}_0$, do not influence the  outcome of the experiment.

The differential evolution algorithm described above generates an offspring population of vectors by addition of a component-wise fluctuation (proportional to the mutation constant $\mu$, see equation \ref{eq:DEA}), whose average amplitude is proportional to the component-wise distance between two members of the parent population. For larger values of $t$ the vectors in a population tend to be confined nearer to the extrema or saddle points of the function  ${\cal L}$; their average distance  tend to decrease and, consequently, also the mutation fluctuations. This feedback loop is such that the distance among population vectors unavoidably tend to decrease, especially in the fast-search regime. The velocity with which the distance among intra-population  vectors decreases is an estimation of the steepness of the function $\cal L$ to be maximised, around its maxima (or of the subject's criterion definiteness, see \ref{sec:experimentalaim}). 

{\bf Alternative way of assessing the subject self-consistency.} In experiment E2 we have estimated  the consistency of different volunteers' criteria,  comparing how close are the populations of vectors sculpted by a given subject in the final generation, $\{\f^{(s,n)}(T)\}_n$, of different realisations of the experiment. An alternative way of ``sampling'' the relevant face-space region of a subject (hence comparing different subjects' relevant regions) could be that of performing a longer experiment in which a proper stationary state is reached, such that the populations in the latest generations of the genetic experiment are statistically indistinguishable. If it exists, the stationary state (under a {\it stationary} dynamical rule), should not depend on the initial condition and on the sequence of numbers. This would require, of course, to modify the algorithm (that otherwise necessarily tends to produce closer and closer generations) with the addition of a fluctuation term that is constant in time: 

\begin{eqnarray}
	v_{i}^{(j)}&=&f^{(j)}_{i}(t) + \mu (f^{(j_1)}_{i}(t)-f^{(j_2)}_{i}(t)) + \chi_i\qquad  \textrm{ with prob. } \rho \\
	v_{i}^{(j)}&=&f^{(j)}_i(t)\qquad \textrm{with prob. } 1-\rho 
\label{eq:DEAalt}
\end{eqnarray}
being $\bm \chi(t)$ a $D$-vector of uncorrelated random numbers (in $t$ and in their component) with null average and small (smaller than $\mu_{\rm sc}$), fixed standard deviation. We propose this variant of the algorithm as an alternative strategy for future experiments.

\section{Calculation of observables and their errors \label{sec:observables}}

The observables that we have calculated to perform the data analysis are functions of the set of sculpted facial vectors, ${\cal S}=\{f_i^{(s,n)}\}_{s=1,n=1}^{S,N}$, defined in a general fashion, independent of the face-space parametrisation (restricted landmark Cartesian coordinates or inter-landmark distances).  They are defined as the following. 
\begin{itemize}
\item {\bf Average of facial vectors.} $\<f_i\>=(1/(SN))\sum_{s,n} f_i^{(s,n)}$.
\item {\bf Standard deviation of facial vectors.} Crucially, the standard deviation of the single facial vector coordinate $\sigma_i$ is not computed along both indices $(s,n)$, ${\rm std}(\{f_i^{(s,n)}\}_{s,n})$, since the facial vectors within the population of a single individual are correlated. As a consequence, the naive standard deviation is an underestimation of the inter-subject standard deviation. The statistical error of the average, $\sigma_{\<f_i\>}$ is calculated instead as a Bootstrap error, or the standard deviation of a series of $B\gg S$ averages of $\f$ over different subsamples ${\cal S}_b$ made by a random, identically uniformly distributed set of $S$ subject indices. For each subject index $s$, only one population facial vector, $n_s$ is considered:

\begin{eqnarray}
	\<f_i\>_b = \frac{1}{S}\sum_{j=1}^S f_i^{(s_j(b),n_j(b))}  \\
\sigma_{\<f_i\>}={\rm std}\left( \<f_i\>_1,\ldots, \<f_i\>_B\right) 
\end{eqnarray}
where, for each $b$, $s_j(b)$, $n_j(b)$ are a set of $S$ independently distributed (in $b$ and in $j$) integer numbers in the intervals $[1:S]$ and $[1:N]$ respectively. In this way, the standard deviation of the average is computed over a set of averages where, in each one, $S$ subjects are used (hence this error is proportional to $\sim S^{-1/2}$, as desired), and such that only uncorrelated (i.e., coming from different subjects) facial vectors are used in each average $\<\cdot\>_b$. The inter-subject error of the single coordinate is then computed simply as:

\begin{equation}
\sigma_i=S^{-1/2}\,\sigma_{\<f_i\>}
\end{equation}

\item  {\bf Standarised variables.} A set ${\cal Y}$ of standarised facial vector is constructed standarising each vector in ${\cal S}$: 

\begin{equation}
y_j^{(s,n)}=(f_j^{(s,n)}-\<f_j\>)/\sigma_j
\end{equation}
\item  {\bf Correlation matrix.} The correlation matrix of a standarised set of vector coordinates ${\cal G}=\{\y^{(s)}\}_{s=1}^N$ is computed as:

\begin{equation}
C_{ij}[{\cal G}]=\frac{1}{S} \sum_{s} y_i^{(s)} y_j^{(s)} 
\label{eq:C}
\end{equation}

The matrix $C$ used to compute the $t$- and $p$- values reported in table \ref{tab:C} and figure \ref{fig:C} is the average and standard deviation of this quantity over a {\it set of populations}. It is computed, again, with the Bootstrap method, in such a way that the sum in eq. \ref{eq:C} runs over different subject populations, and only one facial vector of each population is considered. The Bootstrap error corresponds to the standard deviation from subject to subject, proportional to $S^{-1/2}$. In general, the Bootstrap average and error of an observable ${\cal O}$ of the set of standarised vector populations $\cal Y$ is 

\begin{eqnarray}
\<{\cal O}\> &=& {\rm average}\left( \<{\cal O}\>_1,\ldots, \<{\cal O}\>_B\right) \\
\sigma_{\cal O}&=&{\rm std}\left( \<{\cal O}\>_1,\ldots, \<{\cal O}\>_B\right) 
\end{eqnarray}
where:

\begin{eqnarray}
	\<{\cal O}\>_b = \frac{1}{S}\sum_{j=1}^S {\cal O}[\y^{(s_1(b),n_1(b))},\ldots,\y^{(s_S(b),n_S(b))}] 
\end{eqnarray}
Taking ${\cal O}[{\cal G}]=C[{\cal G}]$ in eq. \ref{eq:C}, we obtain the average and standard deviation of the correlation matrix. 

Since the averages have been subtracted in the standarised variables, the correlation matrix of a set of uncorrelated facial vectors vanishes within their Bootstrap errors, i.e., presents a large $p$-value, $p\sim 1/2$. 

\item  {\bf Principal components.} We define the set of principal components of the facial vectors, ${\cal Y}'=\{{y'}_i^{(s,n)}\}_{s=1,n=1}^{S,N}$, where a vector $\y'$ is the vector of the projections of the vector $\y$ along the various principal axes or eigenvectors of the correlation matrix, $\y'=E \y$ where $E$ is the row-eigenvector matrix, $ECE^{\dag}={\Lambda}$ and $\Lambda={\rm diag}(\lambda_1,\ldots,\lambda_D)$, being ${\lambda_j}$ the $j$-th eigenvalue of $C$.
\item  {\bf Distances between sets of vectors.} Given two sets of vectors, ${\cal S}_1=\{{\bf f}^{(s,n)}\}_{s=1,n=1}^{S_1,N}$, ${\cal S}_2=\{{\bf g}^{(s,n)}\}_{s=1,n=1}^{S_2,N}$, the {\it inter-set pseudo-distance} is defined as:

\begin{equation}
	{\rm dist}({\cal S}_1,{\cal S}_2)=\frac{1}{S_1 S_2}\sum_{s_1,s_2} {\sf D }(\f^{(s_1,\cdot)},{\bf g}^{(s_1,\cdot)}) \label{eq:setdistance}
\end{equation}
		where $\sf D$ is the (per coordinate) inter-population pseudo-distance, defined as:

\begin{equation}
{\sf D }(\f^{(s_1,\cdot)},{\bf g}^{(s_2,\cdot)})=\frac{1}{N^2}\sum_{n_1,n_2} \frac{1}{M} {\sf d} (\f^{(s_1,n_1)},{\bf g}^{(s_2,n_2)})
	\label{eq:popdistance}
\end{equation}
%

%
and where $M$ is the dimension of the vectors ${\bf f}$, $\bf g$, and ${\sf d}({\bf x_1},{\bf x_2})$ is the {\it face-space metrics} or the distance between two single vectors in face-space. It can be defined in various ways (see \cite{hill2011,valentine2016}):

\begin{enumerate}
\item {\bf Euclidean-metrics.} As the Euclidean distance between the principal components of the vectors: ${\sf d}({\f_1},{\f_2})=|| {\f'_1}-{\f'_2} ||$, where $||\cdot||$ is the Euclidean metrics in $D$ dimensions, being $\f'=E\f$, and $E$ the row-eigenvector of the last $r$ eigenvectors of the  correlation matrix (corresponding to non-standarised facial vectors). Taking $r=D$, it coincides with the Euclidean distance, $|| {\f_1}-{\f_2} ||$. 
\item {\bf Euclidean-metrics with standarised vectors.} As the Euclidean distance between the principal components of the standarised vectors: ${\sf d}({\y_1},{\y_2})=|| {\y'_1}-{\y'_2} ||$, being $\y'=E\y$, and $E$ the row-eigenvector of the standarised correlation matrix.
\item {\bf Mahalanobis-metrics.} As the Euclidean metrics between {\it standarised principal components} of the vectors, or: ${\sf d}({\y_1},{\y_2})=|| {\y''_1}-{\y''_2} ||$, where $\y''=\y'/\bm \lambda$, being $\y'=E\y$ (and $\bm \lambda$ the eigenvalues of the correlation matrix), the vector division meaning a component-wise division. 
\item {\bf Angle-metrics.} As the angle (in the $D$-dimensional face-space) subtended between two standarised principal components, or: ${\sf d}({\y_1},{\y_2})={\rm arccos}(\y_1''\cdot\y_2''/||{\y''_1}||\,||{\y''_2}||)$. 
\item {\bf Byatt-Rhodes metrics.} As:

	\begin{equation}	
		{\sf d}({\y_1},{\y_2})=\frac{ || {\y''_1}-{\y''_2} ||\,|\y''_1|\,|\y''_2| } {\y''_1\cdot\y''_2+\epsilon}
	\end{equation}	
$\epsilon$ being a small regularising term.
\end{enumerate}

All the results are quantitatively equivalent using instead the {\it min-inter-population} pseudo distance:

\begin{equation}
	{\sf D }_{\rm min}(\f^{(s_1,\cdot)},{\bf g}^{(s_2,\cdot)})=\frac{1}{2}\frac{1}{N}\left\{\sum_{n_1}\min_{n_2}+\sum_{n_2}\min_{n_1}\right\} \left\{\frac{1}{M} {\sf d} (\f^{(s_1,n_1)},{\bf g}^{(s_2,n_2)})\right\}
\end{equation}

\item {\bf Reducing the number of principal components.} 
		In the main text, we have also analysed the effect of reducing the number of principal components in the definition of the metrics. This is implemented as keeping in $\y'$ (and, consequently, in $\y''$) the principal components of $\y$ corresponding to the $r\le D$ highest eigenvalues of $C$ only. If they are ordered in increasing order,: $\lambda_1\le\lambda_2\le\cdots\le\lambda_D$, this is $\y'=E\y$ being $E$ the $r\times D$ matrix made by the last $r$ row-eigenvectors.

	\item  {\bf Statistical errors of distances.} The statistical error associated to the metrics among sets of populations, equation \ref{eq:setdistance}, is the standard deviation of the argument in the sum across couples of different indices $s_1,s_2$. The statistical error associated to the inter-population pseudo-distance, equation \ref{eq:popdistance}, is the standard deviation of the argument in the sum across couples of different indices $n_1,n_2$. The latter error is lower than the former. 

\end{itemize}

\section {Assessment of the  convergence of populations of vectors}

The degree of coherence of the single subject's criterion in experiment E1 may be estimated through the degree of convergence of the population of vectors sculpted by the subject as a function of the generation index,  $t$. In figure 2 (main article) we show the self-distance between the population of vectors sculpted by 10 randomly chosen subjects as a function of $t$. For a subject $s$, this quantity is (see \ref{eq:setdistance}):

\begin{equation}
d^{(s)}_{\rm conv}(t)={\sf D}(\f^{(s,\cdot)}(t),\f^{(s,\cdot)}(t))
\end{equation}
or the pseudo-distance between the population sculpted by the $s$-th subject at the $t$-th generation and itself. The figure errors have been calculated as explained in the precedent subsection. 

Remarkably, different subjects exhibit different degrees of convergence. The reasons for such a diversity is an argument of possible cognitive interest in itself. A proposal for further work is to investigate the relation among the convergence velocity shown in Fig. 2 (main article), the subject's response times (see \ref{sec:responsetimes}), and the self-consistency distance of each subject (see Fig. 3 (main article)). 

In any case, after some generations, the populations of all the subjects in the sample result more self-distant (see Fig. 2 (main article)) than the populations resulting from a null model of genetic experiment, performed with the same parameters as in experiment E1 but in which the selection  step  (equation \ref{eq:inequality}) is random. In Fig. 2 (main article), the error bars  of the null test self-distance among populations refer to the standard deviation of the self-distance in the $t$-th generation across different realisations of the null model experiment. 

\section{Precision of the experiments \label{sec:precision}}

As previously explained, the vector of facial coordinates $\f$ contains the inter-landmark distance vector $\bf d$ or the reduced set of landmark Cartesian coordinates $\bf L$. In the first case, the coordinates represent distances in units of total facial length. Thus, they are not absolute distances, but proportions. In the second case, the coordinates correspond to pixels, divided by the reference portrait length
in pixels. In both cases, they are floating point quantities, and the systematic error associated to the single vector is limited by the image resolution in pixels, $\gtrsim 400^{-1}$. This is roughly the precision with which we resolve the subject intra-population distance using the Euclidean metrics (per coordinate), i.e., the average distance along the single ``physical'' coordinate, see Fig. \ref{fig:distancesphysical}. The self-consistency distance in physical coordinates is, remarkably, barely twice than the image resolution. The average and standard deviation of self-consistency distances per coordinate using the Euclidean metrics is: $0.0045(9)$. This corresponds to a precision of $0.80(15){\rm mm}$ of the  average female facial length. The intra-subject distances are estimated with an even higher precision (Fig. \ref{fig:distancesphysical}).

\section{Response times \label{sec:responsetimes}}

\begin{figure}[t!]                        
\begin{center} 
\includegraphics[width=.8\columnwidth]{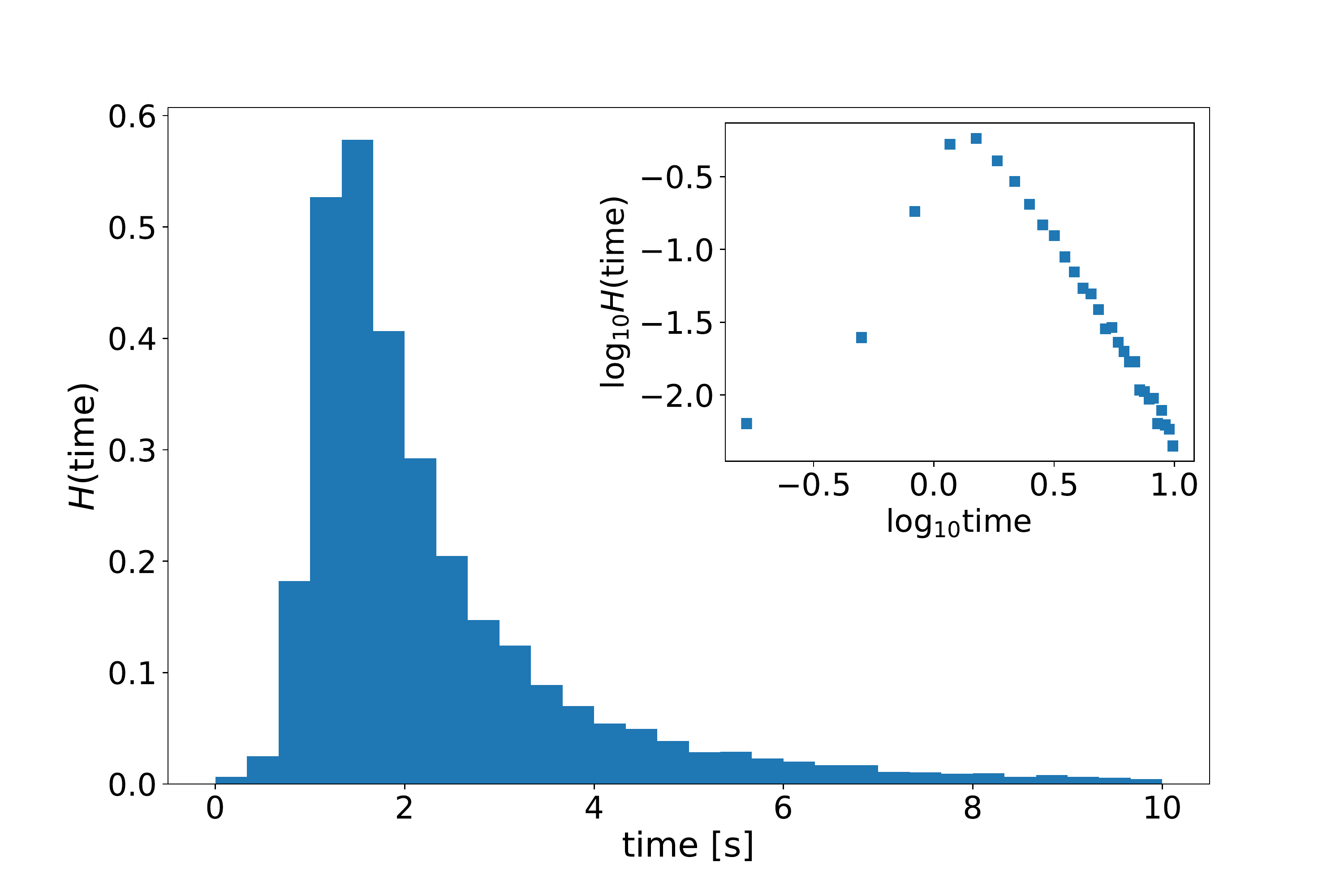}  
	\caption{Histogram of delay time between consecutive left/right choices (between all the 280 choices of all subjects in E1). The most probable time is around $1.75 {\rm s}$. Inset: the histogram in log-log scale.}
\label{fig:responsetimes}
\end{center}   
\end{figure}

As explained in sec. \ref{sec:geneticalgorithm}, the {\bf (3) selection} step of the genetic algorithm is implemented by the human subject, in our experimental scheme, as a choice among two facial images generated by the computer. In figure  \ref{fig:responsetimes} we report the histogram of the ($S_1NT$) elapsed time between consecutive left/right choices of every subject (with $NT=280$ choices for each one) in experiment E1. 

\section{Distances among different  partitions of the dataset \label{sec:distancespartitions}}

\begin{figure}[t!]                        
\begin{center} 
\includegraphics[width=.8\columnwidth]{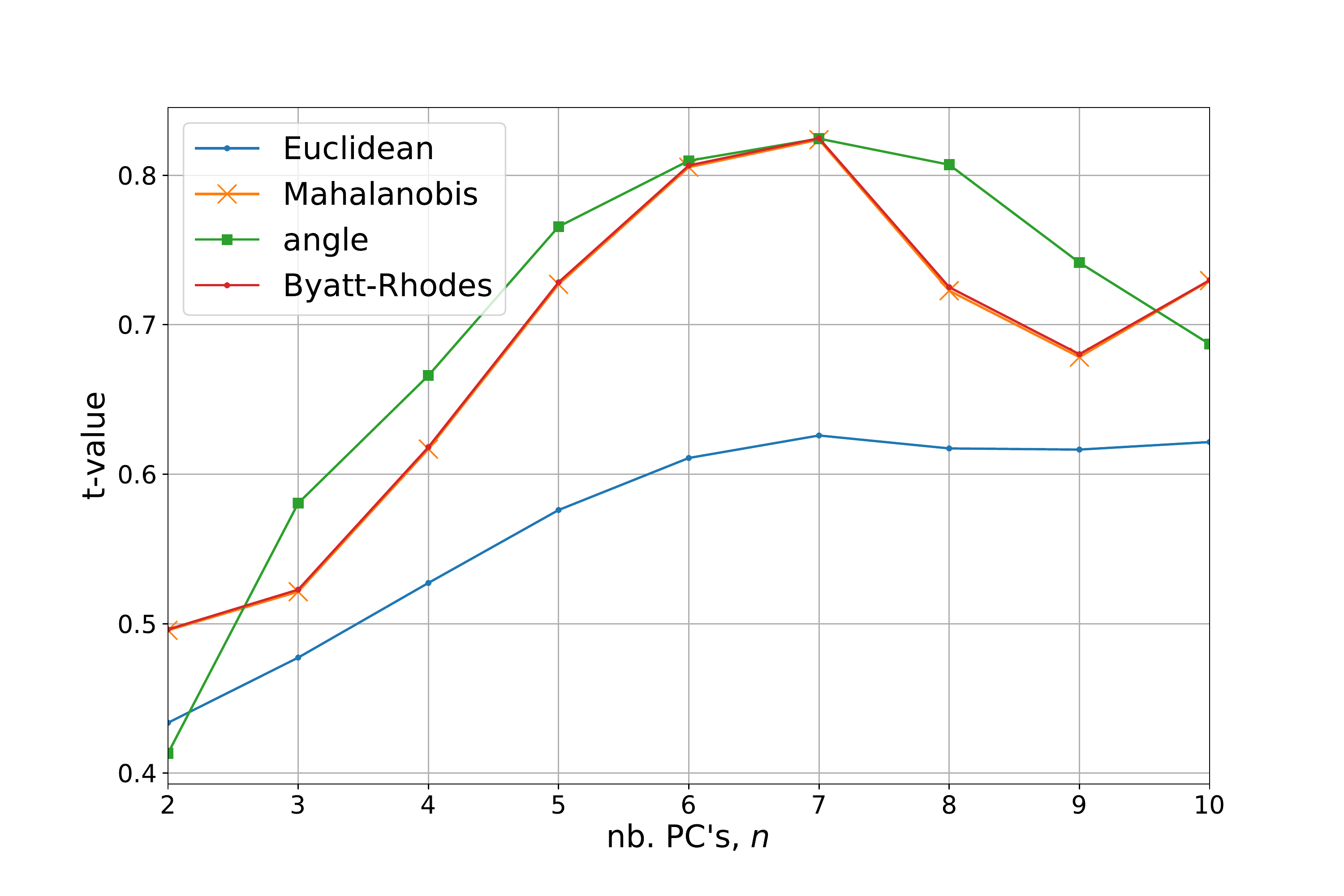}  
	\caption{Histogram difference $t_{12}$ among self-consistent and inter-subject population histograms of distances, versus the number of principal components considered in the definition of the face-space metrics. Different curves correspond to different kinds of face-space metrics.  }
\label{fig:tvaluevariousmetrics}
\end{center}   
\end{figure}

In figure 3 (main article) we report the self-consistency (among couples of populations sculpted by the same subject, for all the subjects in E2) and inter-subject  (among couples of populations sculpted by different subjects in E1) distance histograms. The distances have been computed using the angle-metrics with $n_{\rm pc}=7$ principal components. In figure \ref{fig:tvaluevariousmetrics} we present a comparison of the difference among histograms, $t_{12}=(\mu_1-\mu_2)/(\sigma_1^2+\sigma_2^2)^{1/2}$, using different values of $n_{\rm pc}$ in the face-space metrics definition. Such difference among histograms provides an estimation of the overlap among both distributions: the cumulative normal distribution of $t_{12}$, $p_{12}$, is actually the overlap probability of both histograms, if they are supposed to be normal distributions. The quantity $t_{12}$ is, interestingly, a non-monotone function of $n_{\rm pc}$, the largest self-consistency/inter-subject distance is obtained with the angle-metrics using $n_{\rm pc}=7$ principal components. For this metrics, the probability of two facial vectors sculpted by the same subject to be closer than two facial vectors sculpted by different subjects in E1 is $p_{12}=$\overlappingprobability. This number coincides, within its statistical errors, with the empirical fraction of couples of inter-subject distances that are larger than a self-consistency distance. As mentioned before (see Sec. \ref{sec:geneticalgorithm} for details), this probability is arguably underestimated, due to the finiteness of the experimental procedure.

Notice that $\sigma_i^2$ in the definition of $t_{12}$ are the variances of the histograms, not the variances of the averages of the histograms, $\sigma_i^2/N_i$). These are used to compute the $p$-value of the histogram difference, corresponding to the Student's $t$-value $t=(\mu_1-\mu_2)/(\sigma_1^2/N_1+\sigma_2^2/N_2)^{1/2}$, practically equal to zero, $p<10^{-30}$.


%
 
\begin{figure}[t!]                        
\begin{center} 
\includegraphics[width=.8\columnwidth]{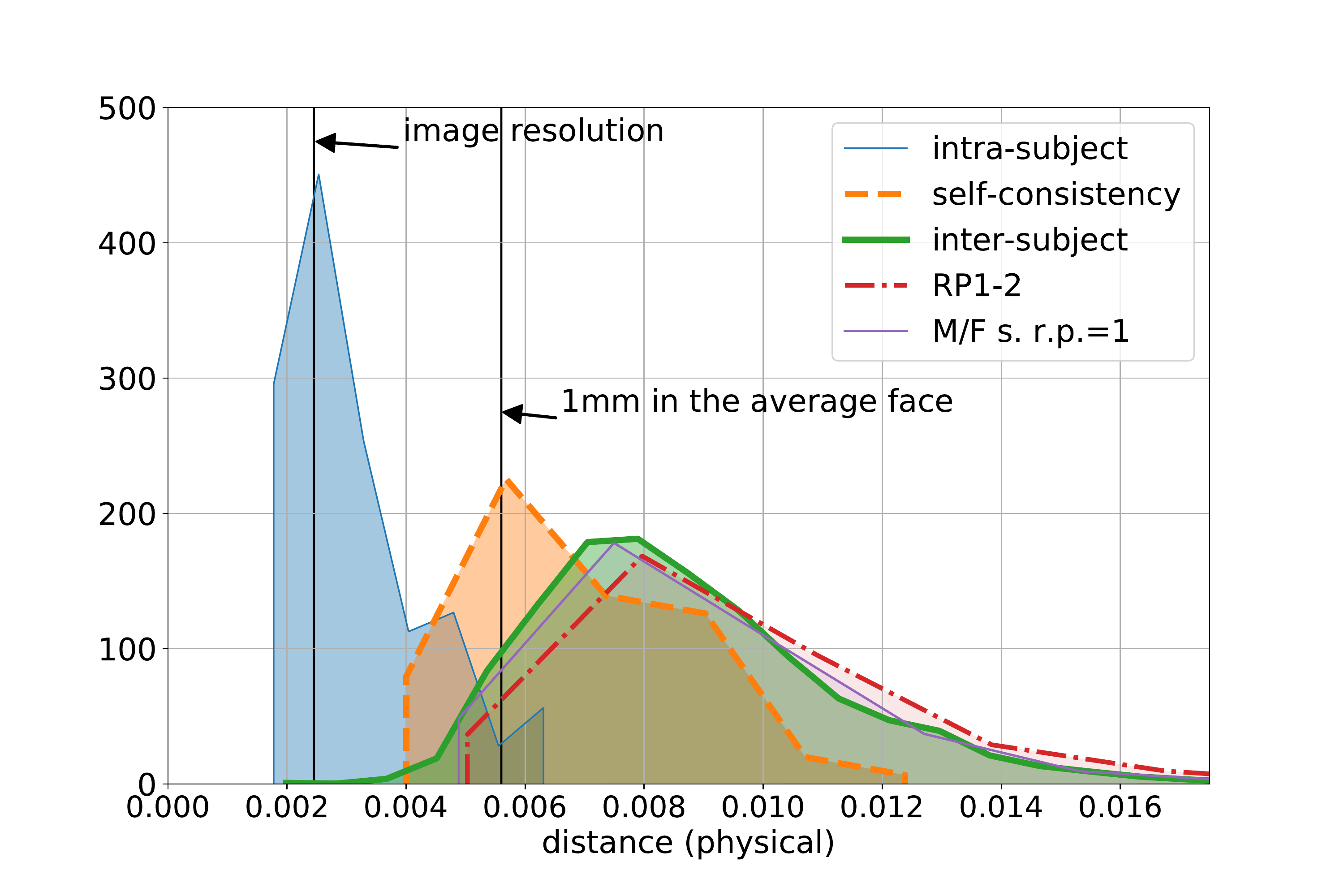}  
	\caption{Histograms of intra-subject, self-consistent and inter-subject sets of inter-population distances (as shown in Fig. 3 (main article) for the angle-metrics), using the Euclidean-metrics in face-space with 11 coordinates (the ``physical'' distances). The arrows indicate the resolution of the image pixel (the distance among two pixels in units of the facial length, $\sim 400^{-1}$), and the scale corresponding to $1{\rm mm}$ ($\ell_{\rm mm}^{-1}$, where $\ell_{\rm mm}$ is the female average facial length measured in $\rm mm$).}
\label{fig:distancesphysical}
\end{center}   
\end{figure}

For completeness, in Fig. \ref{fig:distancesphysical} we present the histograms corresponding to the intra-subject, inter-subject, self-consistent, inter-reference portrait and inter-subject gender sets of inter-population distances calculated with the Euclidean metrics in face-space (see below), i.e. the distance per coordinate using the $D$ physical coordinates. Although the self-consistent and inter-subject histograms are more overlapping  with respect to the angle-metrics histograms, they are still very obviously distinguishable ($p< 10^{-16}$, $p_{12}=0.72(5)$, $t_{12}=0.61(2)$).

Our experimental method succeeds to resolve the differences among single-subject preferred variations. This is possible thanks to the reduction of the face-space dimensionality. Such a reduction is implemented by considering geometrical degrees of freedom only, on the one hand, and, on the other hand, by keeping low the dimension of the (geometric) face-space. 

The ideal number of dimensions is subject to the accuracy/complexity trade-off. A too low-dimensional face-space would not allow to detect the systematic inter-subject differences. As a limit case, think about a face space with a single inter-landmark distance, as the inter-eye distance: in this case, the subjects would clearly not result distinguishable. The differences among different subject’s criteria are more complex, and involve, at least, several (linear combinations of) facial coordinates.  

Conversely, a too high-dimensional face space would not allow a subject to sculpt a consistent (among several realisations) version of his/her attractor in the face-space in a reasonable time. In other words, the resulting set of sculpted faces obtained after a moderate number of choices (of pairwise choices in the software FACEXPLORE), would result less significant, or more dependent on the single realisation of the experiment, and less on the subject’s criterion. 

Such a trade-off is somehow reflected in the non-monotonic behaviour of $t_{12}$ versus $n_{\rm pc}$.

\section{Averages and standard deviations of facial coordinates \label{sec:avstd}}

\begin{figure}[t!]                        
\begin{center} 
\includegraphics[width=.7\columnwidth]{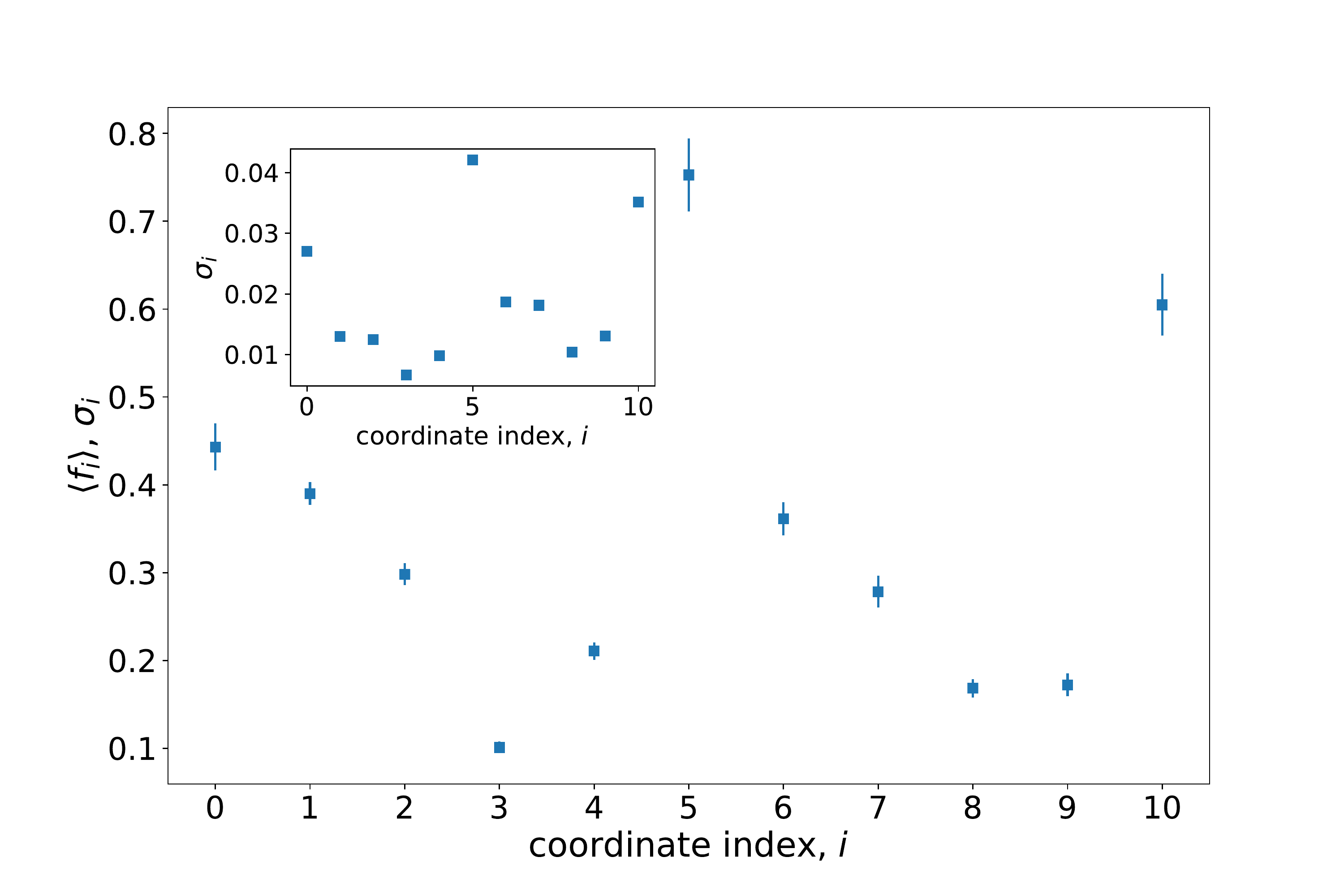}  
	\caption{Averages and standard deviations (as error bars in the main figure, and in the inset) of the experimental facial coordinates, in terms of  inter-landmark distances. }
\label{fig:stds}
\end{center}   
\end{figure}

In figure \ref{fig:stds} we show the results of the experimental averages, $\<f_i\>$ and their standard deviations, $\sigma_i$, in terms of  inter-landmark distances, $f_i=d_i$ (see sec. Methods (main article)). For all the coordinates, the standard deviations are much lower than the averages. 

\section {Statistical distinguishability of partitions of the dataset \label{sec:distinguishability}}

We have performed single component (in physical, $y_i$ or principal, $y'_i$ coordinates) statistical tests between different sets of populations of facial vectors. Given two sets of populations, ${\cal S}_1=\{{\bf f}^{(s,n)}\}_{s=1,n=1}^{S_1,N}$, ${\cal S}_2=\{{\bf g}^{(s,n)}\}_{s=1,n=1}^{S_2,N}$, we consider the $t$-value, and the consequent $p$-value, of the differences among the averages of the $y_i$ (or of the $y'_i$) coordinate in both sets of populations, $\<f_i^{(s_1,\cdot)}-g_i^{(s_2,\cdot)}\>_{s_1,s_2}$. The average $\<\cdots\>_{s_1,s_2}$ is performed again by bootstrapping, summing over  all the couples $(s_1,s_2)$ and over many ($B=500$) realisations in which different population indices $(n_1(s_1),n_2(s_2)$ are chosen for each tuple $s_1,s_2$. In this way, the error of this quantity is of order $\sim (S_1S_2)^{-1/2}$:

\begin{eqnarray}
\<f_i-g_i\>_b	&=& \frac{1}{S_1S_2} \sum_{s_1,s_2=1}^{S_1,S_2} f_i^{(s_1,n_{1,b}(s_1))}-g_i^{(s_2,n_{2,b}(s_2))} \\ 
\<f_i^{(s_1,\cdot)}-g_i^{(s_2,\cdot)}\>_{s_1,s_2} 		&=& {\rm average}\left(\<f_i-g_i\>_1,\cdots,\<f_i-g_i\>_B\right) \\
\sigma({\<f_i^{(s_1,\cdot)}-g_i^{(s_2,\cdot)}\>_{s_1,s_2}})	&=&{\rm std}\left(\<f_i-g_i\>_1,\cdots,\<f_i-g_i\>_B\right)
\end{eqnarray}
where, again, $n_{1,b}(s),n_{2,b}(s)$ are random, uncorrelated (in $s$, in $b$ and in $1,2$) integers in $[1:N]$.

\begin{figure}[t!]                        
\begin{center} 
\includegraphics[width=.7\columnwidth]{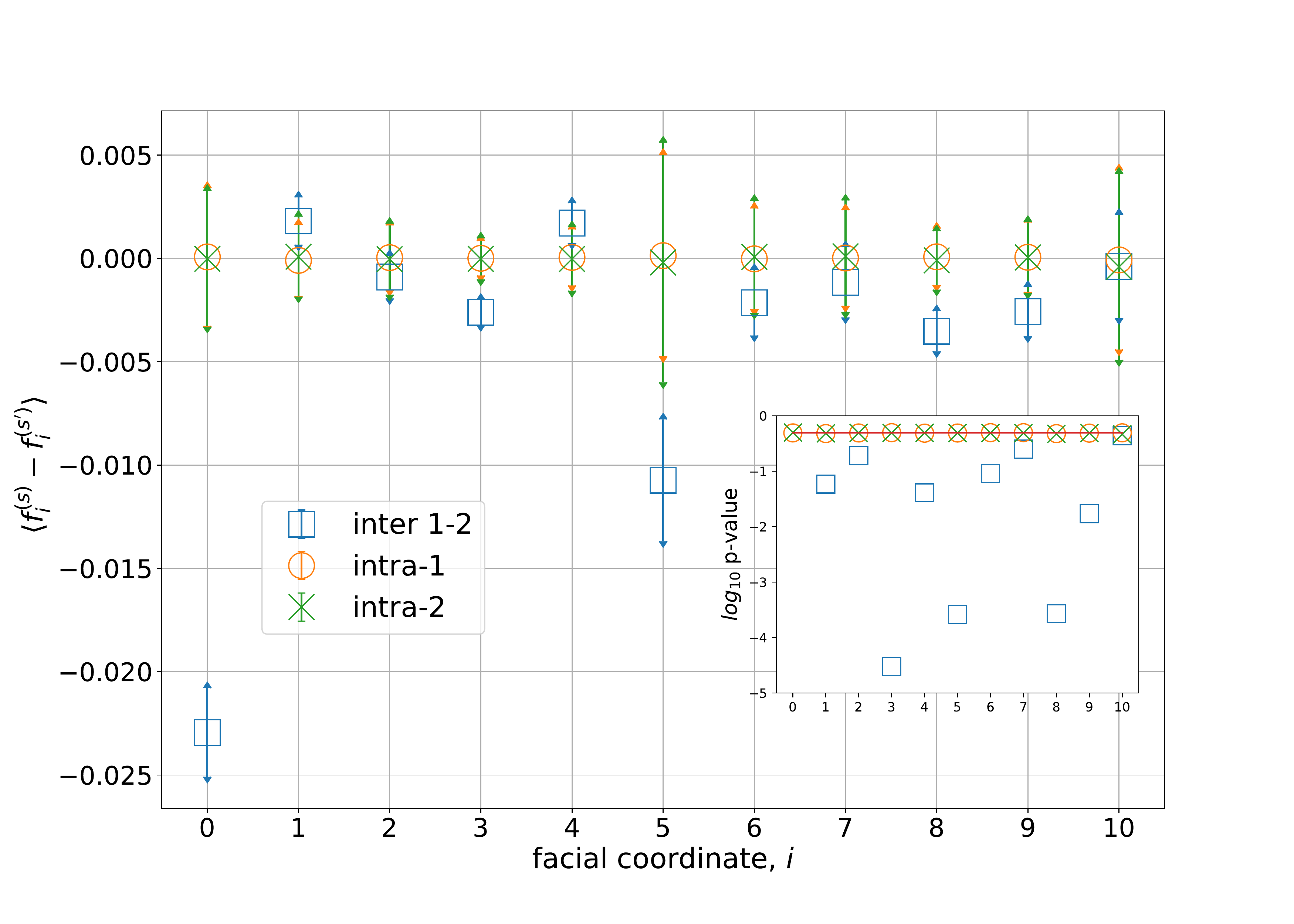}  
\caption{Impact of the reference portrait in different facial coordinates. Differences among facial coordinates $f_i^{(s,\cdot)}-f_i^{(s',\cdot)}$, with $f_i^{(s,\cdot)}$ and $f_i^{(s',\cdot)}$ belonging, respectively, to the set of populations of vectors sculpted in E1 and in E3, as a function of the coordinate index $i$ (squares). Inter-landmark distances have been used as facial coordinates. Circles and crosses are the same quantity, but $f_i^{(s,\cdot)}$ and $f_i^{(s',\cdot)}$ belonging to a random partition of the E1 dataset (circles) and of the E3 dataset (crosses). The error-bars represent the Bootstrap standard deviation, $\sigma(\cdot)$, or the statistical fluctuations of the coordinate differences with respect to the number of subjects only. {\bf Inset}: associated $p$-value (of the $t$-value, $(f_i^{(s,\cdot)}-f_i^{(s',\cdot)})/\sigma(\cdot)$). The $i=2,6,7,10$ coordinates result barely distinguishable or completely undistinguishable.}
\label{fig:12comparison}
\end{center}   
\end{figure}

{\bf According to the reference portrait.} In figure \ref{fig:12comparison} we present the single component differences among the sets of vectors ${\cal S}_1$ and ${\cal S}_2$ described in sec. Results (main article), corresponding to the outcomes of experiments E1 and E3, respectively, in terms of inter-landmark distances as facial vectors, $y_i=d_i$. Only some facial coordinates are distinguishable (within our experimental errors, $\sim \Nv^{-1/2}$) in both sets, specially $d_i$ with $i=3,5,8$. Qualitatively, the same result is found  computing the differences of facial coordinates sculpted by the same subject with different portraits, $f_i^{(s,n_1)}-g_i^{(s,n_2)}$.

\begin{figure}[t!]                        
\begin{center} 
\includegraphics[width=.7\columnwidth]{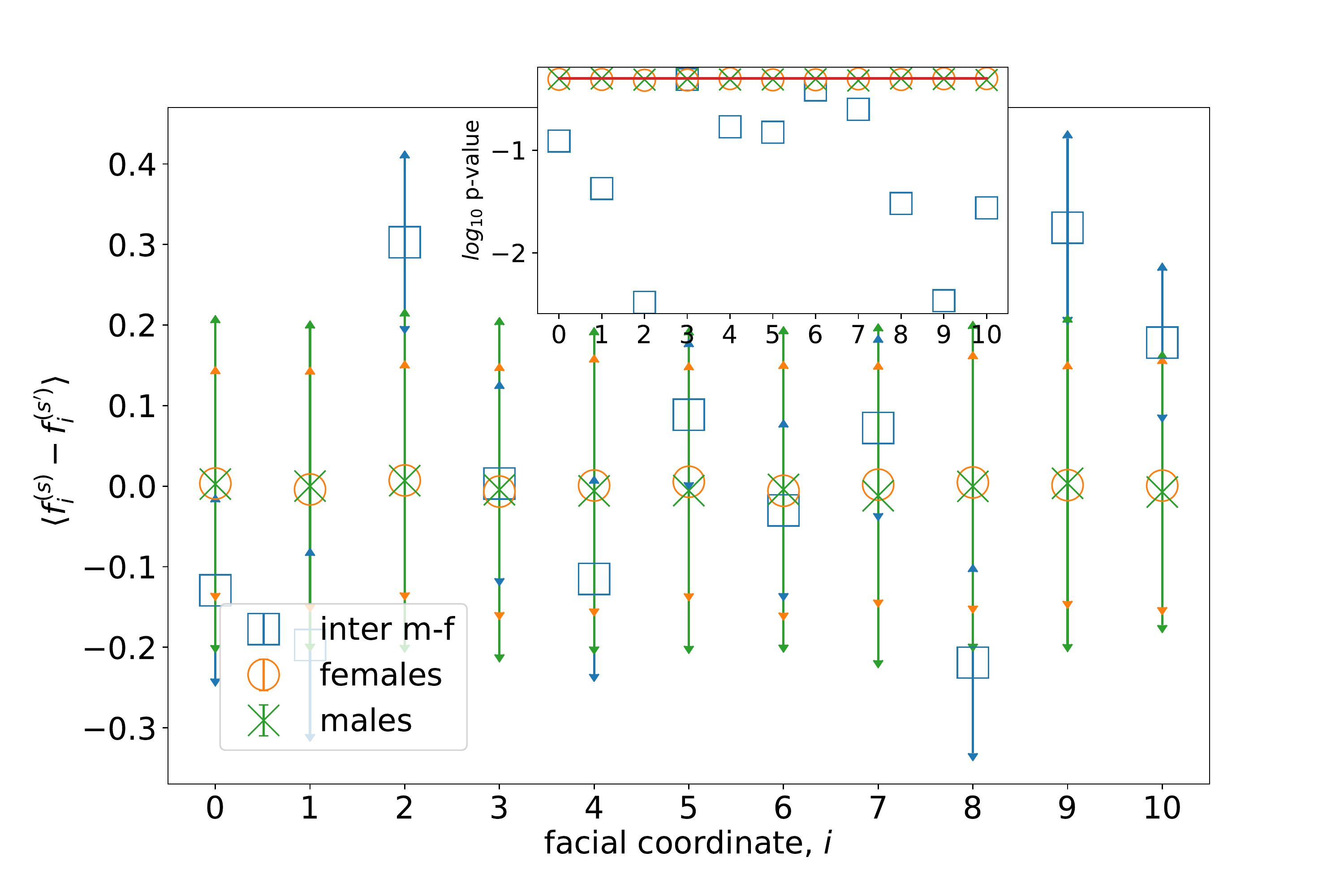}  
\caption{Impact of the subject's gender in different facial coordinates. Differences among facial coordinates $y_i^{(s,\cdot)}-y_i^{(s',\cdot)}$, with $y_i^{(s,\cdot)}$ and $y_i^{(s',\cdot)}$ belonging, respectively, to the set of populations of vectors sculpted by female and subjects in E1, respectively, as a function of the coordinate index $i$ (squares). Inter-landmark distances have been used as facial coordinates. Circles and crosses are the same quantity, but $y_i^{(s,\cdot)}$ and $y_i^{(s',\cdot)}$ belonging to a random partition of the dataset of female (circles) and male subjects (crosses). Symbols, error-bars and inset are as in Fig. \ref{fig:12comparison}. The $i=0,3,4,6,7$ coordinates result barely distinguishable or completely undistinguishable.}
\label{fig:MFcomparison}
\end{center}   
\end{figure}

\begin{figure}[t!]                        
\begin{center} 
\includegraphics[width=.7\columnwidth]{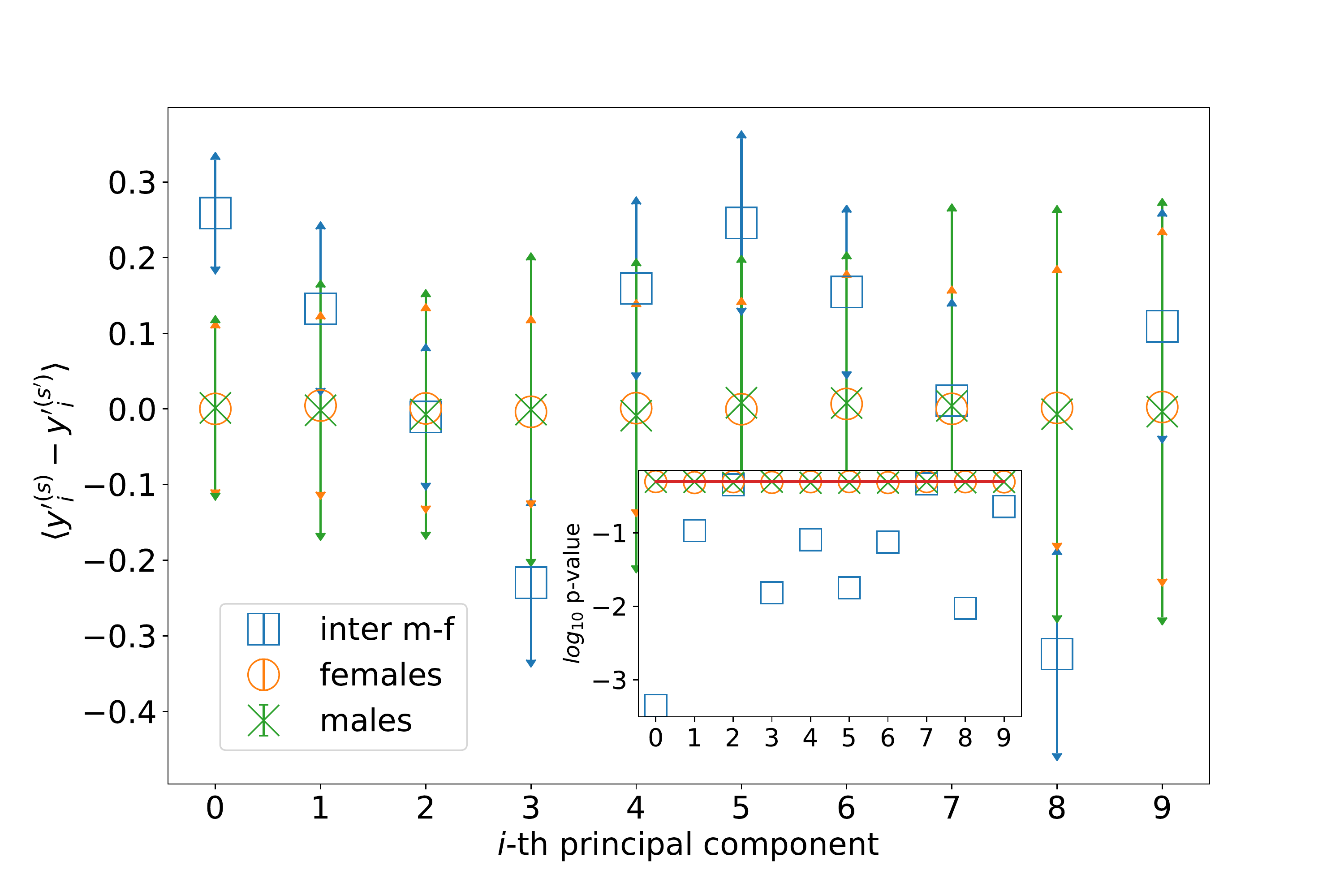}  
\caption{Impact of the subject's gender in different principal components. Differences among principal components of the facial vectors, ${y'}_i^{(s,\cdot)}-{y'}_i^{(s',\cdot)}$, with ${y'}_i^{(s,\cdot)}$ and ${y'}_i^{(s',\cdot)}$ belonging, respectively, to the set of populations of vectors sculpted by female and subjects in E1, respectively, as a function of the coordinate index $i$ (squares). Symbols, error-bars and inset are as in figure \ref{fig:12comparison}. The $i=3,5,7,8,10$ principal components result barely distinguishable or completely undistinguishable.}
\label{fig:MFcomparison_PC}
\end{center}   
\end{figure}

{\bf According to the subject's gender.} In figure \ref{fig:MFcomparison} we present the single component differences among the sets of vectors sculpted by female and male subjects in experiment E1, respectively, in terms of inter-landmark distances as facial vectors, $f_i=d_i$. Only some facial coordinates are distinguishable  in both sets, specially $d_i$ with $i=1,2,8,9,10$ (eye height, nose height, eye width, nose width and zygomatic bone height). In figure \ref{fig:MFcomparison_PC} we show the same results but using the principal components, $y'_i$. Only some principal components ($i=1,4,9$, see subsection \ref{sec:eigenvectorimages} are distinguishable within the experimental errors. The principal component exhibiting largest variability, $y'_{10}$, is barely distinguishable in the female/male subject partition.

\section{Pairwise correlations among facial coordinates \label{sec:correlationmatrix}}

\begin{figure}[t!]                        
\begin{center} 
\includegraphics[width=.5\columnwidth]{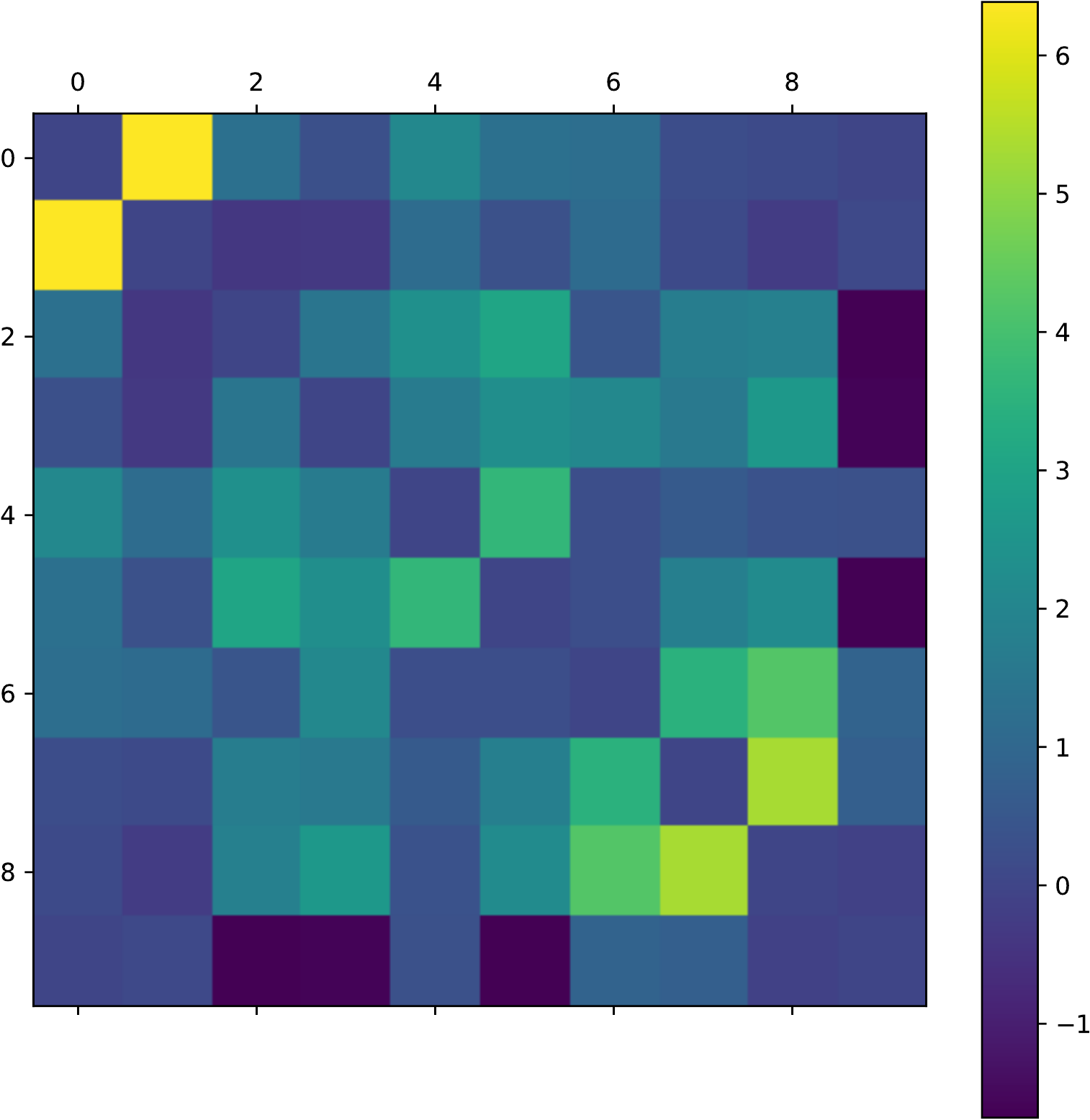}  
\caption{
The $t_{ij}$-value corresponding to the correlation matrix among facial coordinates, $t_{ij}=C_{ij}/\sigma_{C_{ij}}$, in terms of Cartesian landmarks coordinates. The diagonal (equal to the unit vector $C_{ii}=1$) has been set to zero for clarity.
\label{fig:C} }
\end{center}   
\end{figure}

The list of the most strongly interacting couples of facial coordinates (the $C_{ij}$ matrix elements with higher $t_{ij}$-value)  is presented in terms of inter-landmark distances in table \ref{tab:C_dist}, and in terms of landmark Cartesian coordinates in table \ref{tab:C}. Fig. ~\ref{fig:C} presents the $t_{ij}$ matrix elements corresponding to landmark Cartesian coordinates. The error has been calculated as specified in sec. \ref{sec:observables}.

\begin{table}
\begin{tabular}[h]{cllll}
type &  $i,j$  & $|C_{ij}|$ & $t_{ij}=C_{ij}/\sigma_{C_ij}$ & $p$-value \\
\hline
hh & 8,6 & 0.105 & 1.583 & 5.85e-02 \\
hv & 7,0 & 0.102 & 1.613 & 5.51e-02 \\
hv & 9,4 & 0.098 & 1.628 & 5.35e-02 \\
hv & 9,1 & 0.100 & -1.677 & 4.85e-02 \\
hv & 10,0 & 0.104 & 1.722 & 4.42e-02 \\
hv & 10,2 & 0.111 & -1.732 & 4.33e-02 \\
hv & 7,2 & 0.113 & -1.779 & 3.93e-02 \\
hv & 9,0 & 0.104 & 1.807 & 3.70e-02 \\
hv & 8,3 & 0.112 & 1.963 & 2.63e-02 \\
hh & 9,7 & 0.101 & 1.985 & 2.51e-02 \\
hh & 9,5 & 0.104 & 2.054 & 2.14e-02 \\
hh & 9,8 & 0.103 & 2.118 & 1.85e-02 \\
hh & 6,5 & 0.121 & 2.119 & 1.84e-02 \\
hv & 10,3 & 0.109 & 2.133 & 1.78e-02 \\
vv & 3,1 & 0.093 & -2.276 & 1.26e-02 \\
hv & 7,4 & 0.114 & 2.289 & 1.22e-02 \\
hv & 9,3 & 0.108 & 2.497 & 7.16e-03 \\
hh & 7,5 & 0.104 & 2.514 & 6.86e-03 \\
hh & 10,9 & 0.101 & 2.581 & 5.72e-03 \\
vv & 3,2 & 0.117 & -2.643 & 4.84e-03 \\
vv & 4,2 & 0.109 & -2.870 & 2.55e-03 \\
hh & 8,5 & 0.099 & 3.112 & 1.24e-03 \\
hh & 8,7 & 0.113 & 3.242 & 8.31e-04 \\
hh & 10,8 & 0.106 & 3.349 & 5.91e-04 \\
hh & 10,7 & 0.101 & 3.793 & 1.34e-04 \\
vv & 4,1 & 0.122 & -3.872 & 1.02e-04 \\
hh & 10,5 & 0.121 & 3.878 & 9.94e-05 \\
vv & 2,1 & 0.116 & -4.824 & 2.81e-06 
\end{tabular}
\caption{ Relevant experimental correlations $C_{ij}$  in terms of inter-landmark distances along with their corresponding $t$-value.  v/h denotes the vertical/horizontal character of the involved coordinates $d_i$ and $d_j$.  \label{tab:C_dist}}
\end{table}

\begin{table}
\begin{tabular}[h]{llll}
landmarks involved   & $|C_{ij}|$ & $t_{ij}$ & $p$-value\\
\hline
${\sf y}_7 		{\sf x}$ & 0.109 & 1.675 & 4.90e-02 \\
${\sf y}_7 		{\sf x}$ & 0.109 & 1.685 & 4.80e-02 \\
${\sf y}_{14} 		{\sf x}_{10}$ & 0.105 & -1.688 & 4.78e-02 \\
${\sf y}_{14} 		{\sf x}_{12}$ & 0.105 & -1.722 & 4.46e-02 \\
${\sf y}_{7} 		{\sf x}_{7}$ & 0.135 & 1.723 & 4.45e-02 \\
${\sf y}_{9} 		{\sf x}_{7}$ & 0.132 & 1.748 & 4.23e-02 \\
${\sf y}_{9} 		{\sf x}_{12}$ & 0.115 & 2.092 & 1.99e-02 \\
${\sf x}_{14} 		{\sf x}_{1}$ & 0.132 & 2.098 & 1.97e-02 \\
${\sf y}_{1} 		{\sf x}_{10}$ & 0.109 & 2.124 & 1.85e-02 \\
${\sf x}_{14} 		{\sf x}_{7}$ & 0.117 & 2.164 & 1.68e-02 \\
${\sf x}_{12} 		{\sf x}_{10}$ & 0.110 & 2.196 & 1.56e-02 \\
${\sf y}_{9} 		{\sf x}_{10}$ & 0.111 & 2.700 & 4.28e-03 \\
${\sf x}_{12} 		{\sf x}_{7}$ & 0.113 & 3.003 & 1.82e-03 \\
${\sf x}_{12} 		{\sf x}_{14}$ & 0.142 & 3.245 & 8.79e-04 \\
${\sf y}_{7} 		{\sf y}_{1}$ & 0.129 & 3.513 & 3.77e-04 \\
${\sf y}_{9} 		{\sf y}_{1}$ & 0.124 & 4.234 & 3.22e-05 \\
${\sf y}_{9} 		{\sf y}_{7}$ & 0.170 & 5.018 & 1.70e-06 \\
${\sf x}_{3} 		{\sf x}_{1}$ & 0.136 & 6.495 & 4.03e-09
\end{tabular}
	\caption{Relevant experimental correlations $C_{ij}$ in terms of landmark Cartesian coordinates, along with their corresponding $t$-value, $t_{ij}=C_{ij}/\sigma_{C_{ij}}$.  The first column indicates ${\sf c(i)}_{\alpha(i)}{\sf c(j)}_{\alpha(j)}$, where $\alpha(i)$ is the landmark index involved and ${\sf c}(i)={\sf x} $ or $\sf y$.  \label{tab:C}}
\end{table}

\section{Image deformations along principal axes \label{sec:eigenvectorimages}}

\begin{figure}[t!]                        
\begin{center}
\begin{tabular}[h]{lc}
	$i$ & ${\cal I}[{\bf y}_i(\eta)]$, $\eta=-3q,-q/2,0,q/2,3q$
 \\
\hline
$1$ & \includegraphics[width=.6\columnwidth]{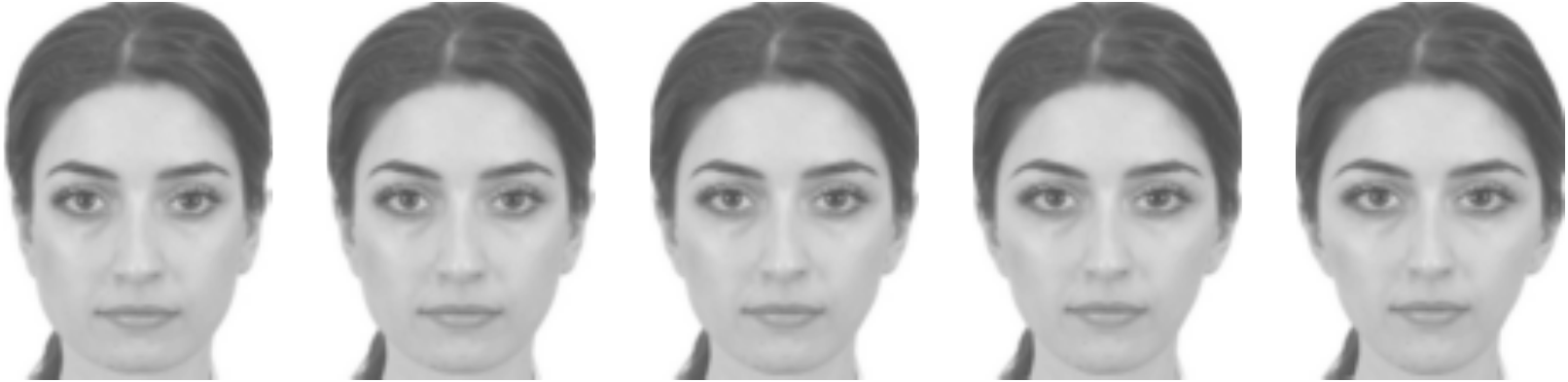}  \\
$2$ & \includegraphics[width=.6\columnwidth]{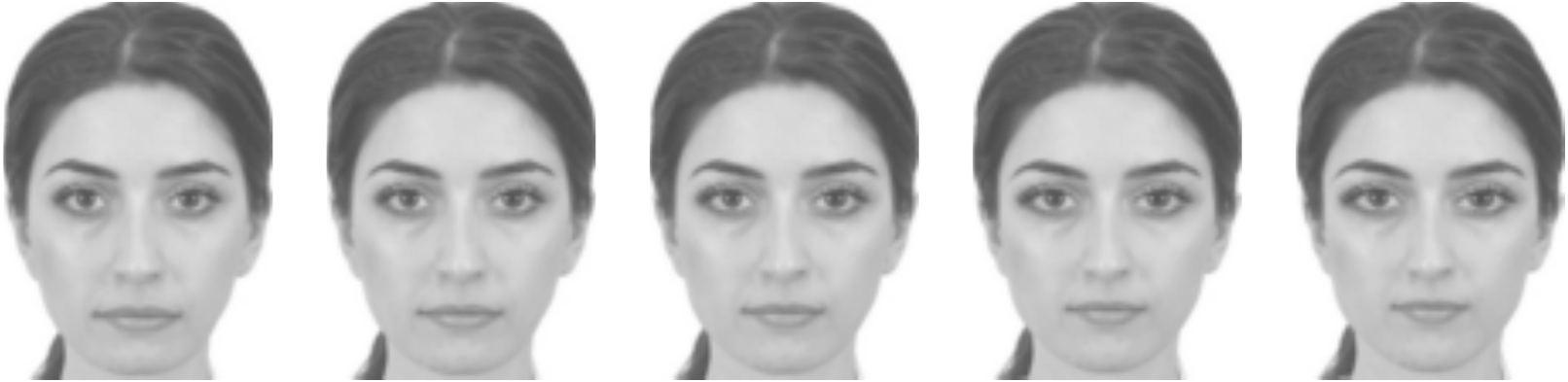}  \\
$3$ & \includegraphics[width=.6\columnwidth]{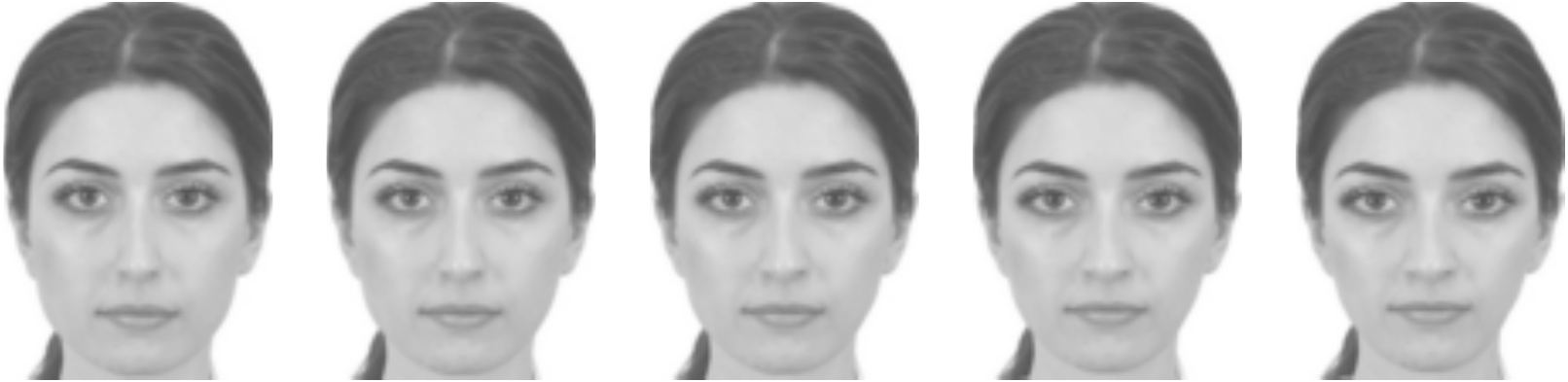}  \\
$4$ & \includegraphics[width=.6\columnwidth]{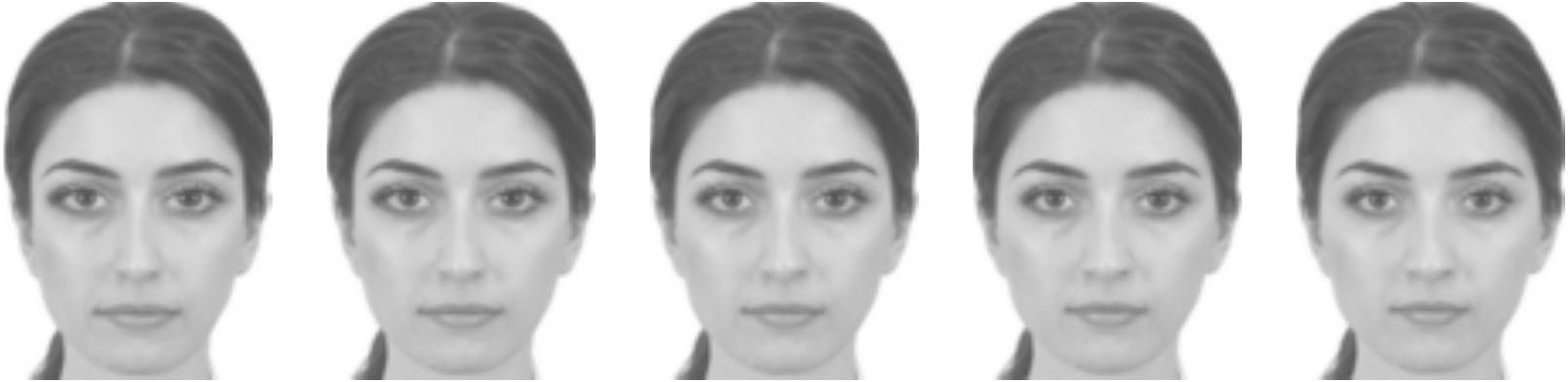}  \\
$5$ & \includegraphics[width=.6\columnwidth]{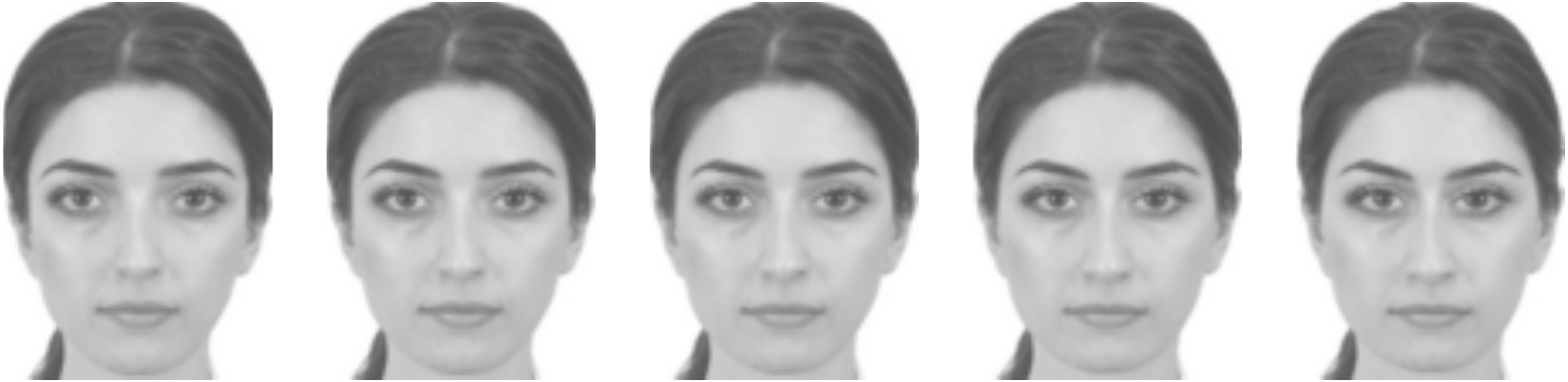}  \\
$6$ & \includegraphics[width=.6\columnwidth]{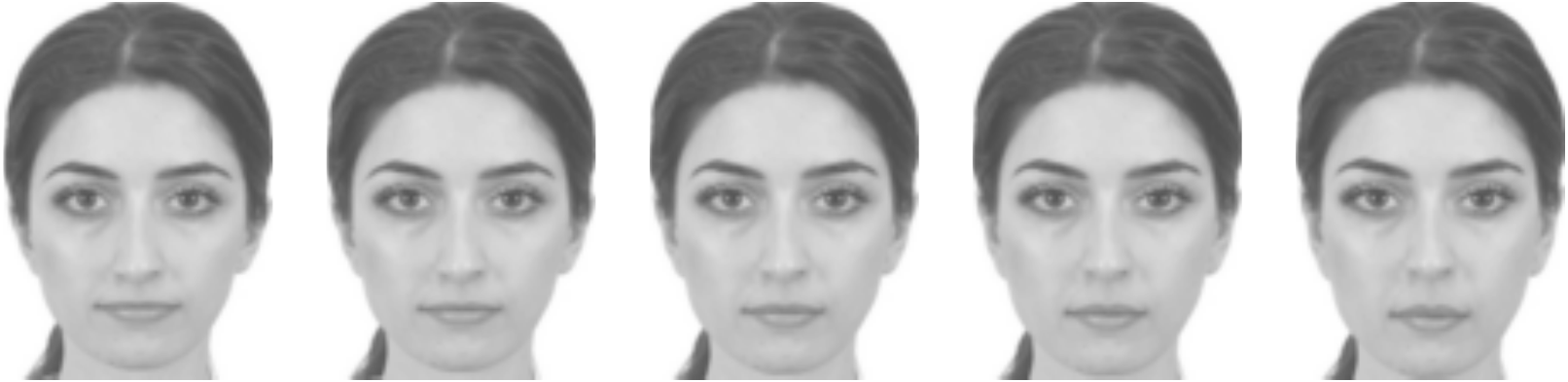}  \\
$7$ & \includegraphics[width=.6\columnwidth]{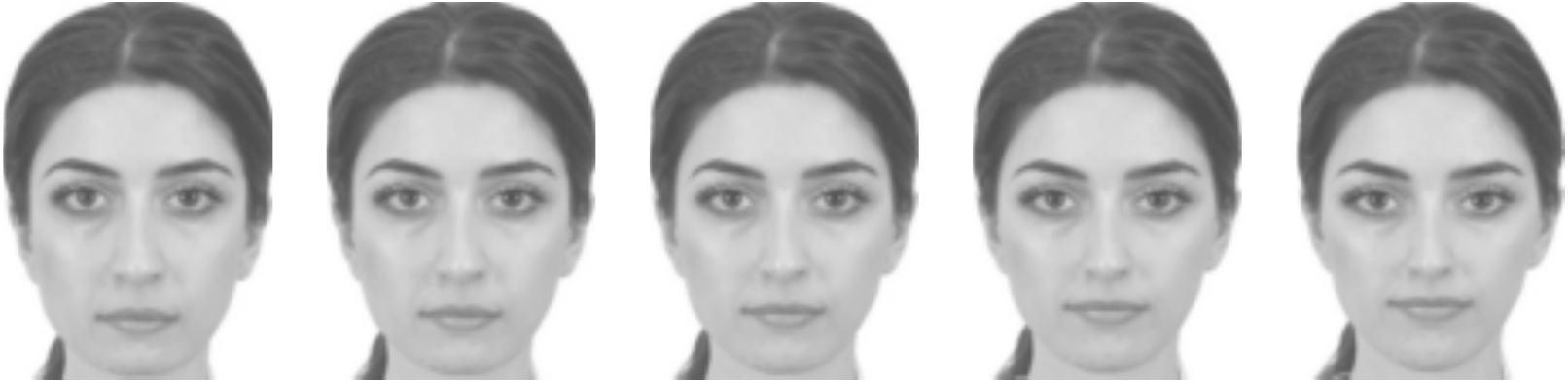}  \\
$8$ & \includegraphics[width=.6\columnwidth]{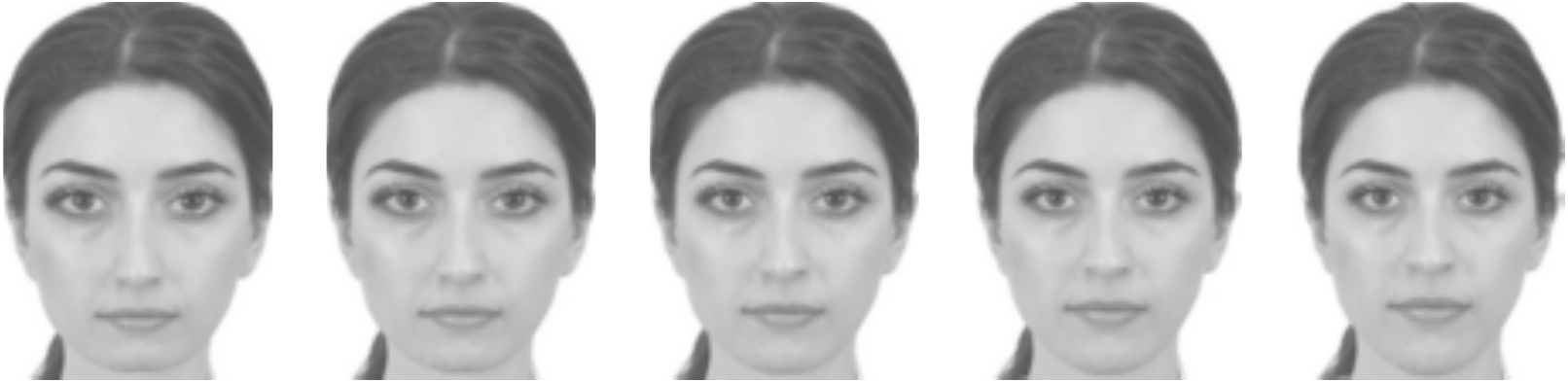}  \\
$9$ & \includegraphics[width=.6\columnwidth]{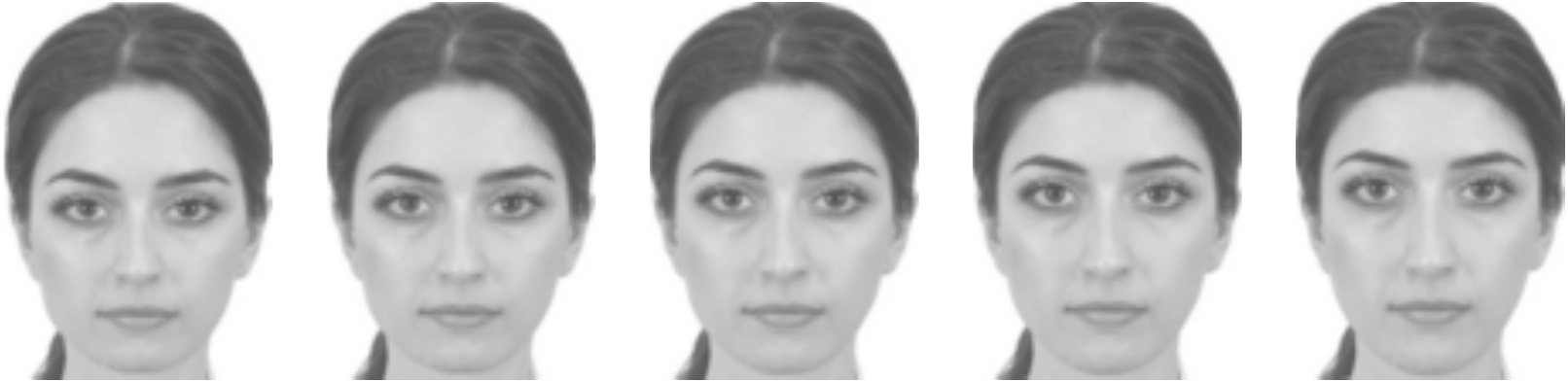}  \\
$10$ & \includegraphics[width=.6\columnwidth]{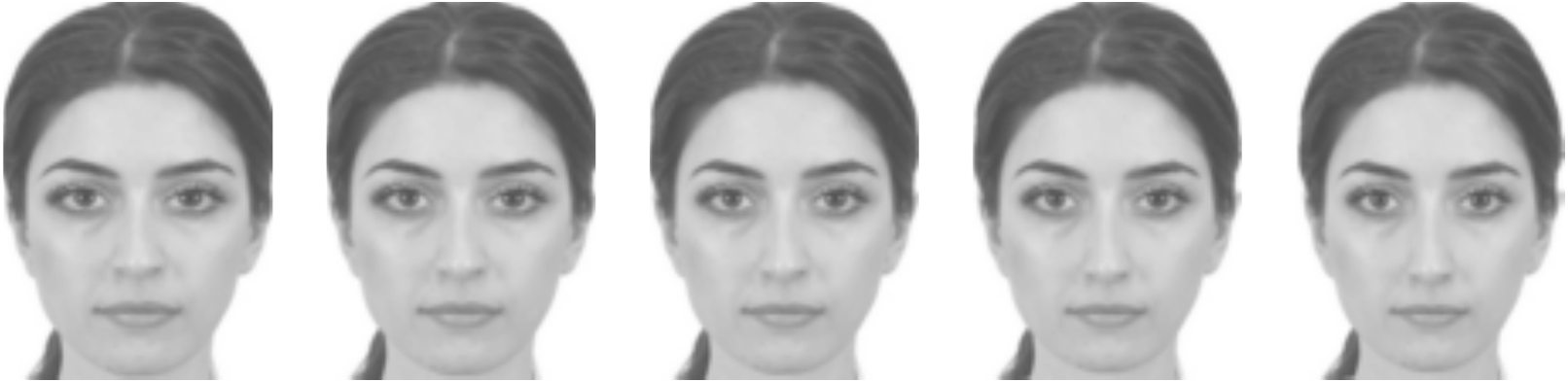}  
 \end{tabular}
	\caption{Facial images ${\cal I}[\eta\, {\bf e}^{(i)}]$ corresponding to various eigenvectors $i$ (in different rows). Different columns correspond to various values of $\eta$, the central column ($\eta=0$) is the average sculpted facial image in all rows. While high rows represent uncommon deformations (at a given $\eta$) with low $C$-eigenvalue $\lambda_i$, lower columns represent common deformations expanding most of the dataset variability.  }
\label{fig:eigenvectorsall}
\end{center}
\end{figure}

\begin{figure}[t!]                        
\begin{center} 
\includegraphics[width=.7\columnwidth]{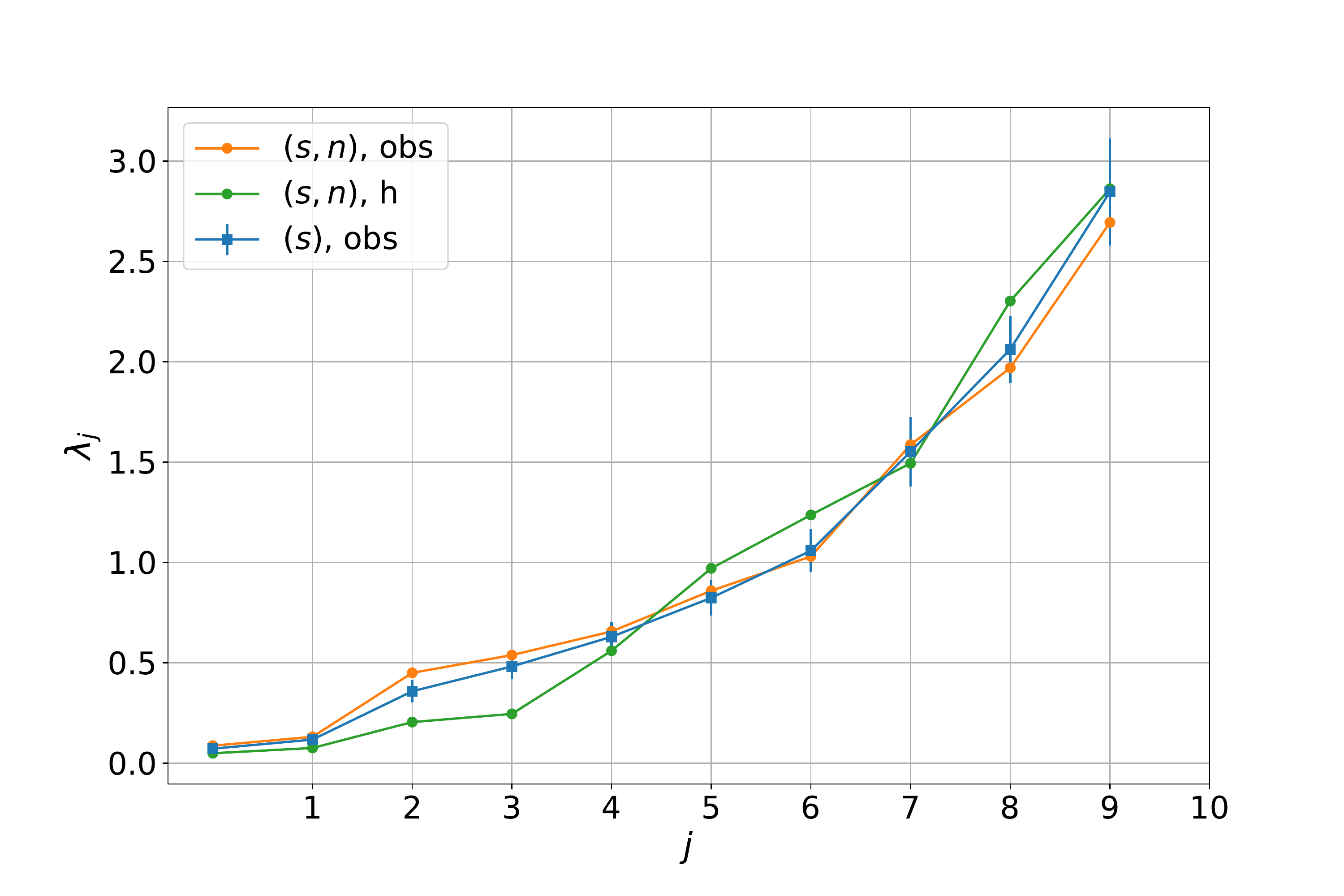}  
\caption{Spectrum of matrix $C$ (blue curve). Larger than one eigenvalues correspond to eigenvectors that vary more than the physical (standarised) components, and vice-versa. The orange curve is the spectrum of the correlation matrix obtained as the average of $y^{(s,n)}_iy^{(s,n)}_j$ over both subject and population indices, $(s,n)$. The green line is the spectrum of matrix $\Ch$ (see \ref{sec:higherorder}). }
\label{fig:spectrum}
\end{center}   
\end{figure}

\begin{figure}[t!]                        
\begin{center} 
\includegraphics[width=.7\columnwidth]{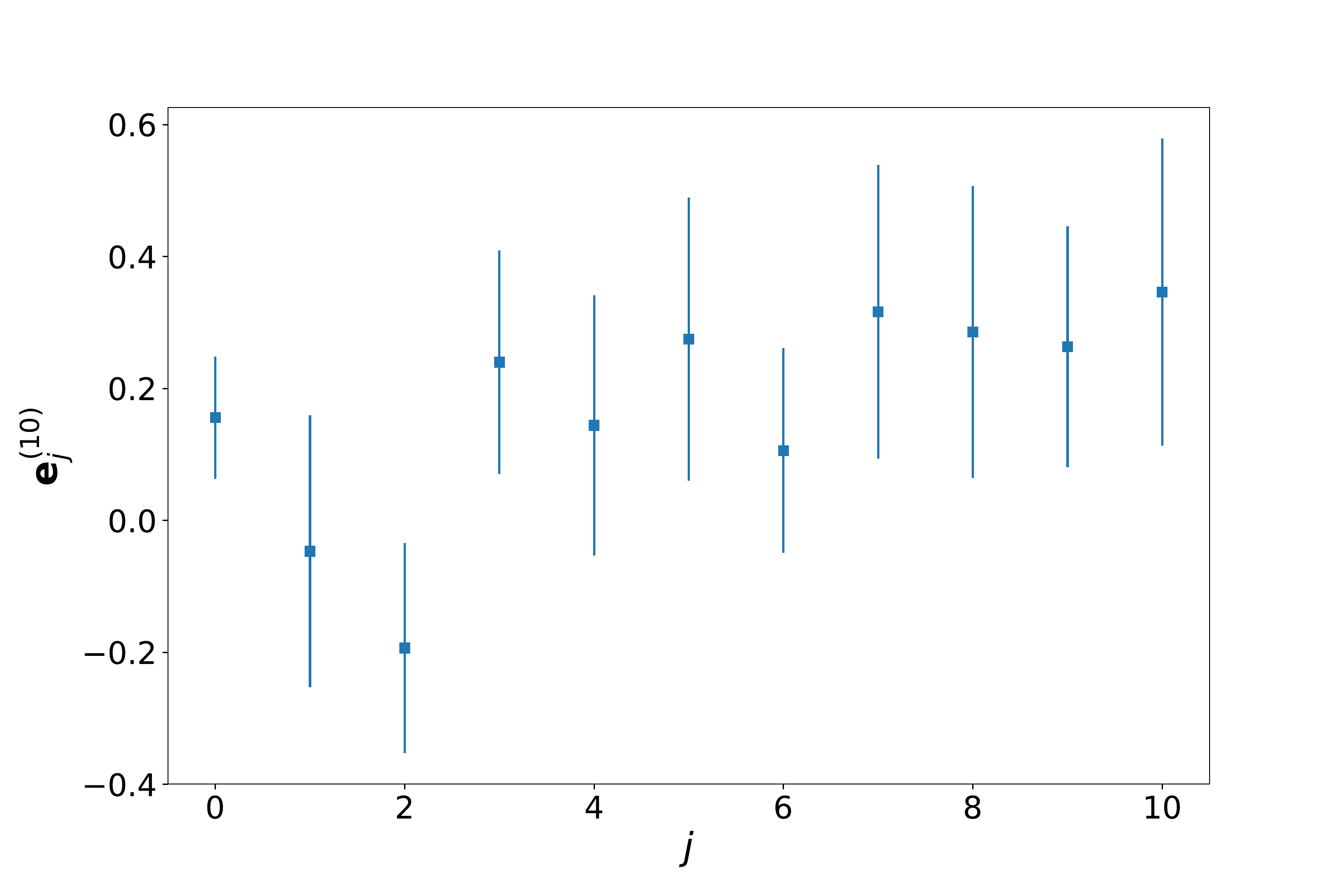}  
\caption{The vector components of the eigenvector ${\bf e}^{(10)}$ of matrix $C$ (for inter-landmark distances), $e^{(10)}_j$, vs. $j$. The error-bars have been calculated by bootstrapping. The $10$-th eigenvector (see figure \ref{fig:eigenvectorsall}) is essentially a scale transformation of the horizontal quantities (barely by the same factor, except for the inter-eye distance), and a linear combination of vertical quantities with smaller factors. }
\label{fig:eigenvectorcomponents}
\end{center}   
\end{figure}

In figure \ref{fig:eigenvectorsall} we report the facial images ${\cal I}[{\bf y}(i,\eta)]$ corresponding to the vectors:

\begin{eqnarray}
{\bf y}(i,\eta) = {\bf 0} + \eta\, {\bf e}^{(i)} \nonumber
\end{eqnarray}
i.e., to a deformation of the average facial vector along the $i$-th principal axis, or the $i$-th  eigenvector of $C$ (from inter-landmark distances), in increasing order of eigenvalues, $\lambda_1<\lambda_2<\cdots<\lambda_{D-1}$ (the $0$-th eigenvalue is null and correspond to the constraint ${\sf h}=1$). Every row of figure \ref{fig:eigenvectorsall} corresponds to a different eigenvector, while each column corresponds to a value of $\eta$ in the set $-3q$, $-q/2$, $0$, $q/2$, $3q$, $q=0.075$ (the central column corresponding to the average facial vector). Higher rows, corresponding to lower associated eigenvalues $\lambda_i$, represent uncommon (say, unpleasant) deformations with respect to other axes (for equal $\eta$'s), since their associated standard deviation, $\lambda_i^{1/2}$, is lower (c.f. figure \ref{fig:spectrum}). The ${\bf e}^{(1)}$ eigenvector, for instance, is a linear combination such that the face width increases while the jaw width decreases, or vice-versa, an uncommon, ``forbidden'' deformation according to the correlation matrix (see table \ref{tab:C_dist}), which indicates that these distances tend to increase or decrease together. Conversely, the lower rows, corresponding to high eigenvalues (larger than one, i.e., larger than the standard deviation of the single coordinates $y_i$), represent common deformations. The last eigenvector, $i=10$, consists mainly in deformations in which the horizontal distances positively covary, by a roughly common, positive factor (see the ${\bf e}^{(10)}$ eigenvector components in figure \ref{fig:eigenvectorcomponents}). The eigenvectors $i=1,4,9$ are the ones along which male and female subjects are more distinguishable. 

\section{Relevant angles \label{sec:relevantangles}}

In sec. Results (main article) we have presented a geometric interpretation of {\it the sign} of the oblique correlation matrix elements in terms of some {\it relevant inter-landmark segment angles}, shown in figure 4 (main article). The sign of a given oblique matrix element (say, in terms of Cartesian landmark coordinates) $C_{ij}=\<\delta{\sf y}_{\alpha(i)}\delta{\sf x}_{\alpha(j)}\>$ (with ${\delta}{\sf x}_\alpha={\sf x}_\alpha-\<{\sf x}_\alpha\>$, and the same for $\sf y$) coincides with that of the slope of the average inter-landmark line, ${\Delta}_{\sf y}/{\Delta}_{\sf x}$, where $\vec{\Delta}=\<\vec\ell_\alpha\>-\<\vec\ell_\beta\>$. This is the only way in which the fluctuation of the  $\alpha$--$\beta$ segment slope around its average value, $(\Delta_{\sf y}+{\delta{\sf y}_{\alpha}})/(\Delta_{\sf x}+{\delta{\sf x}_{\beta}}) - \Delta_{\sf y}/\Delta_{\sf x}$ may vanish, at first order in the $\delta$'s. 

This provides a clear interpretation of matrix $C$: the fluctuations generated by various subject's different aesthetic criteria are such that they tend to respect some {\it natural angles of the face, those defined by the inter-landmark segments evidenced in figure 4 (main article)}. The figure shows the inter-landmark segments corresponding to the oblique correlations $\<{\sf y}_\alpha \sf x_{\beta}\>$ {\it exhibiting higher $t$-value}: perhaps representing the  most relevant angles. One could be tempted, at this point, to attribute a {\it quantitative estimation of the relative importance}  to each one of these segments, proportional to the $C$ $t$-value or modulus. The following arguments suggest that this method is not the optimal way of assessing such relative relevance of various inter-landmark segments, and provide a further motivation to the application of the Maximum Entropy method to this problem. 

%

A rigorous assessment of the relevance of various inter-landmark angles or slopes cannot be directly addressed from matrix $C$, since a slope, in the general case, is defined as a correlation among four landmark Cartesian (or inter-landmark distance) coordinates ($({\sf y}_{\alpha}-{\sf y}_{\beta})/({\sf x}_{\alpha}-{\sf x}_{\beta})$). Furthermore, the empirical correlations are, in principle, an indirect manifestation of the {\it effective interactions} which cause them. The Maximum Entropy method allows to {\it infer} such effective interactions, from which we can more significantly assess the relative importance of various inter-landmark segments \cite{ibanezberganza2019}. 

As can be seen in table \ref{tab:C} and figure 4 (main article), the sign of $C_{ij}$ coincides with that of $\Delta_{\sf y}/\Delta_{\sf x}$ for all the oblique $C_{ij}$ elements, except $\<{\sf y}_{9} {\sf x}_{10}\>$. This matrix element has a reason to be peculiar: the $9$-th and $10$-th landmarks are subject to a constraint. In the construction of the facial image,  ${\sf y}_9-{\sf y}_{10}$ is constant for all the vectors in the dataset (and equal to the value of this quantity in the reference portrait). 
For a similar reason, we have not included the segment $7$--$12$ in figure 4 (main article), despite there being a strong oblique correlation involving both quantities, since this correlation reflects another {\it a priori} constraint in the dataset: the $12$-th landmark height coincides with the mouth's height by construction. ${\sf x}_{12}={\sf x}_7={\sf x}_{18}$. Equivalently, in the case of inter-landmark distances (see table \ref{tab:C_dist}), the presence of the constraint ${\sf h=1}$ induces a null eigenvalue in $C$ and leads to negative correlations among some facial coordinates $d_{1,2,3,4}$, not directly interpretable, as with $\<{\sf y}_{9}{\sf x}_{10}\>$. The Maximum Entropy method can provide an interpretation of the results of the empirical matrix $C$, correcting the artifact induced by such constraints (or quasi-constraints) \cite{ibanezberganza2019}.

To summarise, the inter-landmark segment angles provide a clear interpretation of $C$, which, in its turn provide a cue of the most relevant angles that we evaluate when we form impressions about a face. To perform a rigorous assessment of the relative importance of various facial elements, however, one should use an inference technique going beyond the bare correlation.

\section{Application of the Maximum Entropy method\label{sec:maximumentropy}}

The above arguments motivate a Maximum Entropy-based approach to the problem \cite{janes1957,nguyen2017}. The goal is to infer a probability distribution ${\cal L}(\y)$ from the experimental dataset $\cal S$ (see \ref{sec:experimentalaim}). $\cal L$  reproduces by construction some data sufficient statistics (at least two-coordinate correlations). ${\cal L}(\y)$ is a probabilistic generative model of the dataset $\cal S$, and can be interpreted as the probability of the facial image with facial coordinates $\y$ (and fixed reference portrait) of being sculpted by any subject (or by a given subject having selected or sculpted the dataset $\cal S$). Inferring ${\cal L}$, one also infers a matrix (or a tensor) of {\it effective interactions} between couples (or $p>2$-plets) of facial coordinates, that reflect the relative influence of the facial feature ins each other. This approach provides a theoretical framework allowing to rigorously account for {\it a priori} correlations and constraints, and to address on information-theoretical grounds the relative relevance of various variables \cite{ibanezberganza2019}. 


\section{Higher order and spurious correlations\label{sec:higherorder}}

A natural and relevant question is to what extent higher-order correlations of the data are statistically significant. In other words, whether three-coordinate empirical correlations  $\<y_iy_jy_k\>$ in our dataset exhibit a large $t$-value. The answer is that, although we do observe non-negligible three-coordinate correlations, we cannot attribute a cognitive origin to them--they are rather generated by an artifact of the genetic algorithm. In experiments E1-3, the vectors of the initial population, $\f^{(s,n)}(0)$ (see sec. \ref{sec:geneticalgorithm}) exhibit small 2- and 3-distance correlations that self-propagate and grow through the generations. Indeed, different populations sculpted (after $T=10$ generations) by the genetic null model (with random left-right choices) exhibit significant 2- and 3-distance correlations, $C^{(2,3)}_{\rm null}$. The null correlations are to be ``subtracted'' from those of the experiments with human subjects, $C^{(2,3)}_{\rm obs}$, in order to isolate relevant correlations of cognitive origin only ($C^{(2,3)}_{\rm h}$). This is an interesting inference problem {\it per se}, of wide generality. It would arise also in experiments with natural facial images that are selected by subjects. The $\Cn$ correlations in this case would correspond to the background correlations corresponding to the database of natural images, prior to the selection by the subjects. 

\begin{figure}[t!]                        
\begin{center} 
\includegraphics[width=.7\columnwidth]{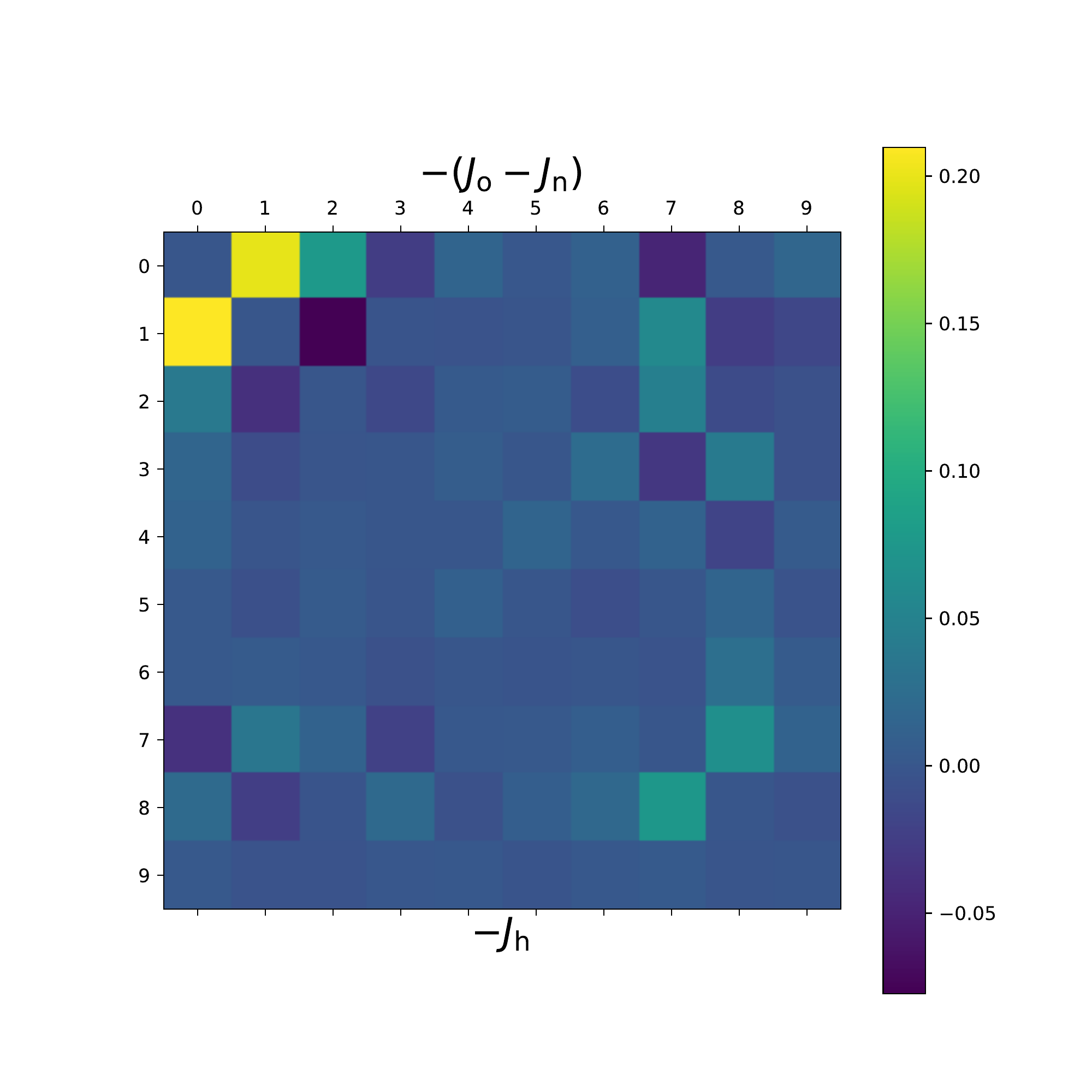}  
\caption{Comparison among the matrices $\Jo-\Jn$ and $\Jh$ for Cartesian landmark coordinates. The upper triangle corresponds to the $-(\Jo-\Jn)$ matrix elements (the minus sign allows for a direct comparison with the $C$ matrix). The lower triangle, to $-\Jh$ matrix elements. The diagonal has been set to zero for a clearer comparison. }
\label{fig:gabrielli}
\end{center}   
\end{figure}

{\bf Subtraction of 3-order null correlations.} On the one hand, 3-order correlation tensors $C^{(3)}_{\rm obs}$ and $C^{(3)}_{\rm null}$ coincide within their statistical errors. We interpret this fact concluding that our experimental scheme does not allow to elicit relevant 3-coordinate correlations of cognitive origin. Nevertheless, we believe that such high-order correlations may exist and play a role in the cognitive process of facial discrimination, and would probably emerge in larger experimental datasets, with higher values of $S$. {\bf Subtraction of 2-order null correlations.} On the other hand, the 2-coordinate correlation matrix $C^{(2)}_{\rm obs}$ measured from the experimental data of E1,3 is clearly distinguishable from $C^{(2)}_{\rm null}$, and exhibits larger matrix elements in absolute value. The ``subtraction'' of null from observed correlations cannot be performed simply as $C^{(2)}_{\rm h}=C^{(2)}_{\rm obs}-C^{(2)}_{\rm null}$, since this leads to a non-positive definite matrix in general.  An alternative method, that we will motivate, analyse and describe in detail in a forthcoming communication, is given by the following procedure: (i) the non-standarised connected correlation matrices corresponding to the null experiment and to the experiments with humans are first defined: ${\tilde \Cn}_{ij}=\<f_if_j\>-\<f_i\>\<f_j\>$ and so with $\tilde\Co$; (ii) one then defines the {\it interaction matrices}: $\Jo=\tilde\Co^{-1}$, $\Jn=\tilde\Cn^{-1}$; (iii) the interaction matrix $\Jh$ is defined in the following way:

\begin{eqnarray}
\Jh={E_{\rm obs}}^\dag \, {\rm diag}({\bm \epsilon_{\rm h}})\, E_{\rm obs} \\
	{\epsilon_{\rm h}}_i ={\epsilon_{\rm obs}}_i - \left[  E_{\rm obs} \Jn  {E_{\rm obs}}^\dag \right]_{ii}  \label{eq:Jh} 
\end{eqnarray}
where $E_{\rm obs}$ is the matrix diagonalising $\Jo$ (or $\tilde\Co$) and $\bm \epsilon_{\rm h}$ are its eigenvalues; (iv) one defines $\Ch=\Jh^{-1}$; (v) finally, one standarises the matrix $\Ch$: $C_{ij}=\Ch_{ij}\,({\epsilon_{\rm h}}_i{\epsilon_{\rm h}}_j)^{1/2}$. In steps (i-v) we have just lowered each eigenvalue of matrix $\Jo$, ${\epsilon_{\rm obs}}_j$, by a quantity which is the expected value of $\Jn$ according to the corresponding eigenvector of $\Jo$. Since the effective null {\it interaction matrix} (in the language of Maximum Entropy inference) is much lower than the interactions of cognitive order, the $D$ quantities ${\epsilon_{\rm h}}_i$ are all positive. We have used this method to isolate the spurious and artifact correlations $C^{(2)}_{\rm null}$ from the observed experimental correlations $\Co$ in E1,E3, leading to the matrix called $C$ throughout the article.  The $C$ matrix elements so obtained are very close to that of the matrix $\Co-\Cn$ (but the matrix $C$ is positive definite). In Fig.~\ref{fig:gabrielli} we show a comparison of $\Jo-\Jn$ vs. $\Jh$, for which this comparison is more evident since these matrices are not subject to the standarisation constraint. This fact suggests that the method efficiently ``removes'' the spurious, artifact correlations induced by the genetic algorithm from the data. A definite confirmation will be provided in future experiments, in which the correlations present in the initial condition of the genetic algorithm will be removed. In any case, {\it all the results} presented in this article are qualitatively {\it identical} using simply $C=\Co$. In the future publication \cite{ibanezberganza2019} we will present a further rigorous method to ``subtract'' $\Cn$ from $\Co$, motivated in the context of the Maximum Entropy method. 

\section{Bibliography}

\printbibliography[heading=none]